\documentclass[pshowpacs,aps,prb,floatfix,twocolumn]{revtex4}
\usepackage{epsfig}

\newcommand{\R}{{\bf r}}

\begin{document}


\title{Correlation Effects on Magnetic
Anisotropy in Fe and Ni }
\author{Imseok Yang}
\email{iyang@physics.rutgers.edu}
\affiliation{Department of
Physics and Astronomy and Center for Condensed Matter Theory,
Rutgers University, Piscataway, NJ 08854}

\author{Sergej Y. Savrasov}
\email{serguei.savrassov@njit.edu}
\homepage{http://physics.njit.edu/~savrasov}
\affiliation{Department of Physics, New Jersey Institute of
Technology, Newark, NJ 07102}

\author{Gabriel Kotliar}
\email{kotliar@physics.rutgers.edu}
\homepage{http://www.physics.rutgers.edu/~kotliar}
\affiliation{
Department of Physics and Astronomy and Center for Condensed
Matter Theory, Rutgers University, Piscataway, NJ 08854}
\date{\today}

\begin{abstract}
We calculate magnetic anisotropy energy of Fe and Ni by taking
into account the effects of strong electronic correlations,
spin-orbit coupling, and non-collinearity of intra--atomic
magnetization. The LDA+U method is used and its equivalence to
dynamical mean--field theory in the static limit is derived. The
effects of strong correlations are studied along several paths in
$(U,J)$ parameter space.  Both experimental magnitude of MAE and
direction of magnetization are predicted correctly near $U=1.9\
eV$, $J=1.2\ eV$ for Ni and $U=1.2\ eV$, $J=0.8\ eV$ for Fe. The
modified one--electron spectra by strong correlations are
emphasized in conjunction with magnetic anisotropy.
\end{abstract}

\maketitle


\section{Introduction}

One of the long-standing problems that are still short of detailed
understanding is to explain the magneto-crystalline anisotropy
energy of magnetic materials containing transition-metal elements,
especially that of Fe, Co, and Ni. Magnetic anisotropy is the
dependency of internal energy on the direction of spontaneous
magnetization.  Generally, the magnetic anisotropy energy term
possesses the symmetry of the crystal. It is therefore called
magneto-crystalline anisotropy or crystal magnetic anisotropy. In
transition metals, most of the magnetic moment comes from spin
polarization. Within the non-relativistic approach, there is no
term coupling to the spin degrees of freedom. The magnetic
moment, therefore, can points to an arbitrary direction. The
magnetic anisotropy energy is an relativistic phenomenon of
spin-orbit coupling. With spin-orbit coupling, the spin degrees
of freedom interact with the spatial anisotropy through the
coupling to the orbital degrees of freedom. This induces a
preferred direction of spins. This happens even though the total
angular momentum is completely quenched. Since spin-orbit
coupling couples individual spin degrees of freedom to individual
orbital degrees of freedom, the fact that total orbital magnetic
moment is quenched does not affect the visibility of spatial
anisotropy to magnetic moment.

The primary difficulty toward investigating MAE has been
attributed to the fact that the MAE of the metals are as small as
of the order of $1\ \mu eV/$atom. At low temperature ($T=4.2 K$),
MAE is of the order of $60\ \mu e V/$atom for hcp Co, $2.8 \mu e
V/$atom for fcc Ni, and $1.4\ \mu eV/$atom for bcc Fe. With
advances in the accurate total energy method combined with the
development of faster computers, attempts have been made to
calculate MAE from first-principles for bulk crystalline Fe, Co,
and Ni. While the magnitude of the MAE have been predicted for all
the three metals, the correct easy axis for Ni has not been
predicted so far. Here we present new results that predict the
correct easy axis for Ni as well as reproduce the previous results
in relevant limits. This result is obtained by incorporating
spin--orbit coupling, non-collinear nature of of intra--atomic
magnetization, and strong correlation effects. We also suggest why
the previous approaches have failed in obtaining the correct easy
axis for Ni.

Experimentally, there are various ways of measuring the magnetic
anisotropy energy.  These range from torque measurements,
saturation magnetic fields and ferromagnetic resonance methods.
However, torque measurements are more often used and are
considered to be reasonably precise. The main principle behind
the torque  experiments is that in the presence of an external
magnetic field misaligned with respect to the easy axis (or
equivalently,
 direction of the crystal magnetization), the magnetic moment experiences a torque.
 In equilibrium,
this torque  is balanced by the crystal anisotropy torque. Torque
magneto--meters are used to precisely measure this torque.
 Since the crystal anisotropy torque is just the angular derivative of the magnetic
anisotropy energy, a measurement of the torque gives an indirect
estimate of the anisotropy energy. The torque measured at various
angles are interpolated to a corresponding analytic expression
such as angular derivatives of total energy. The analytic
expression of total energy can be expressed in terms of
parameters based on crystal symmetry.

 The
simplest form of crystal magnetic anisotropy is uniaxial
anisotropy, for example, in hexagonal cobalt with easy direction
parallel to the $c$ axis of the crystal at room temperature.  As
the internal magnetization rotates away from the $c$ axis, the
anisotropy energy increases with increase of $\phi$, where $\phi$
is the angle between the $c$ axis and the internal magnetization.
We can expand this energy in a series of powers of $\sin^2 \phi$:
   \begin{equation}
      E_a = K_1 \sin^2 \phi +K_2 \sin^4 \phi +\cdots,\label{ex1}
   \end{equation}
where $K_1$ and $K_2$ are constants.

For cubic crystals such as iron and nickel, the anisotropy energy
can be expressed in terms of the directional cosines $\left(
\alpha_1, \alpha_2, \alpha_3 \right)$ of the internal
magnetization with respect to the three cubic edges. Because of
the high symmetry of the cubic crystal, the anisotropy energy can
be expressed in a fairly simple way:
   \begin{equation}
      E_a = K_1 \left( \alpha_1^2 \alpha_2^2
                       +\alpha_2^2 \alpha_3^3
                       +\alpha_3^2 \alpha_1^2 \right)
          + K_2 \left( \alpha_1^2 \alpha_2^2 \alpha_3^2 \right)
          + \cdots,\label{ex2}
   \end{equation}
where $K_1$ and $K_2$ are constants.

  The constants $K_n$ obtained by
interpolating the torque are substituted back to get the
expression for the total energy. Magnetic anisotropy energy is
calculated by subtracting energies at two different direction of
magnetic moments, e.g. [001] and [111]

To measure the torque, specimens are formed into short cylinders.
The specimens are mounted in a carriage held by two torsion fibers
that are fastened to a rigid support at the top and to a circular
scale S at the bottom. When the field is excited in the
electromagnet, the crystal tends to turn so that the direction of
easy magnetization is parallel to the field. The torque so
produced is balanced by turning the bottom of the lower fiber
until the crystal regains its original orientation as determined
by reflection of a light beam from the mirror. The scale reading
$S_2$ is then compared with the original reading $S_1$ with $H=0$.
$S_2-S_1$ is a measure of the torque. The orientation of the
crystal axes with respect to the applied field is varied by
turning the electromagnet, which is mounted on a heavy bearing,
and noting its position on a suitable scale, $S'$. One then plots
the torque against the crystal orientation and deduces from the
curve the crystal anisotropy constant $K_n$.

Early attempts to explain magnetic anisotropy are based on the
interaction between the magnetization and the lattice through
spin-orbit coupling combined with band theory~\cite{vanVleck:37}.
Later, Brooks~\cite{brooks} used an itinerant electron model and
orbital angular momentum quenching in cubic crystals to explain
MAE. Treating the spin-orbit coupling as a perturbation, the
nontrivial magnetic anisotropy came out at the  fourth-order. The
correct directions of magnetization were obtained for Fe and Ni.
The difference between easy axes in these metals was attributed to
the different lattice structures of them. Based on   Ni-Fe alloy
data, however, a much closer correlation between the anisotropy
and the number of valence electrons was observed. What is the
factor that determines the easy axis has been another unanswered
question since then.  In subsequent
papers~\cite{fletcher,sloncewskij,asdente} the calculations became
more and more refined. Finally the discussion centered around the
importance of degenerate states along symmetry lines in the
Brillouine zone and plausible explanations of the origin of
magnetic anisotropy were provided~\cite{Kondorskii,mori:74}.

Contemporary studies of magnetic anisotropy energy have centered
on first-principles calculations, using density functional
theory~\cite{hk64} within the framework of local spin density
approximation (LSDA)~\cite{lsda}.  Eckardt, Fritsche, and
Noffke~\cite{efn} were able to get the values of the right order
of magnitude for Fe and Ni, but with incorrect easy axis for Fe.
Daalderop, Kelly, and Schuurmans~\cite{dks} used the force
theorem~\cite{ma} to obtain the correct order of magnitude of the
MAE for Fe, Co, and Ni, but incorrect easy axes were found for
Co, and Ni. They also observed that changing the number of valence
electrons, hence removing the LDA artificial $X_2$ pocket would
restore the correct easy axis. This observation is similar to
Brooks' observation of the close correlation bewteen the direction
of magnetic moment and the number of valence electrons. Wang, Wu,
and Freeman developed state tracking method to study magnetic
anisotropy of Fe~\cite{freeman}. Trygg, Johansson, Eriksson, and
Willis~\cite{tjew} improved the method while using fully
self-consistent approach. Orbital polarization was also
incorporated, which was suggested by Jansen~\cite{jansen}. While
the correct easy axis was obtained for Co and Fe, in case of Ni
the calculation still gave the wrong easy axis. This is the best
result before the current work. Schneider, Erickson, and
Jansen~\cite{sej} used torque instead of energy difference to
obtain the same result as that of Trygg, Johansson, Eriksson, and
Willis. They also treated the spin-orbit coupling constant as an
adjustable parameter. They succeeded in restoring the correct
easy axis for Ni with an unphysically large value of spin-orbit
coupling. This work suggested a close relation between magnetic
anisotropy energy and the strength of spin-orbit coupling.
Halilov, Perlov, Oppeneer, Yaresko ,and Antonov also scaled
spin-orbit coupling in order to enlarge its effect on the MAE and
confirmed previous results~\cite{halilov}.

 We believe that the physics of transition metal compounds
is intermediate between atomic limit where the localized $d$
electrons are treated in the real space and fully itinerant limit
when the electrons are described by band theory in k space. A
many--body method incorporating these two important limits is the
dynamical mean--field theory (DMFT)~\cite{gabi:rmp}. The DMFT
approach has been extensively used to study model Hamiltonian of
correlated electron systems in the weak, strong and intermediate
coupling regimes. It has been very successful in describing the
physics of realistic systems, like the transition metal oxides
and, therefore, is expected to treat properly the materials with
$d$ or $f$ electrons.

We take a new view that the correlation effects within the $d$
shell are important for the magnetic anisotropy of  $3d$
transition metals like Ni. These effects are not captured by the
LDA but are described by Hubbard--like interactions presented in
these systems and need to be treated by an extension of  first
principles methods such as DMFT. Since, DMFT reduces to
LDA$+$U~\cite{anisimov} in static limit, we adopt LDA$+$U method
to attack the problem of magnetic anisotropy of $3d$ transition
metals.

Another effect which has not been  investigated  in the context of
magnetic anisotropy calculations is the non-collinear nature of
intra-atomic magnetization~\cite{singh}. It is expected to be
important when spin-orbit coupling and correlation effects come
into play together. We show that when we include these new
ingredients into the calculation we solve the long-standing
problem of predicting the correct easy axis of  Ni. Part of this
work has been published elsewhere~\cite{imseok}.

The remainder of this paper is organized as follows.
Section~\ref{ldaudmftsec} presents DMFT extended to relativistic
electron structure problem, its equivalence to LDA+U in static
limit. Section~\ref{maesec} describes calculated MAE.
Section~\ref{bandsec} discusses the calculated MAE in conjunction
with band structure. Section \ref{soosec} presents how to include
other relativistic corrections than spin--orbit coupling and
discuss the effects of spin--other--orbit coupling. Section
\ref{concsec} is a summary and conclusion.

 \section{DMFT and LDA$+$U} \label{ldaudmftsec}

 The LDA
method is very successful in many materials for which the
one--electron model of solids works. However, in correlated
electron system this is not always the case.  In strongly
correlated situations, the total energy is not very sensitive to
the potential since the electrons are localized due to the
interactions themselves, and the lack of sensitivity of the
functional to the density, does not permit to devise good
approximations to the exact functional in this regime. For
example, when the Mott transition takes place the invertibility
condition is not satisfied. Our view is that these difficulties
cannot be remedied by using more complicated exchange and
correlation functionals in density functional theory. DMFT is a
method successfully describing strongly correlated
systems~\cite{gabi:rmp} and has been extended to electronic
structure problems~\cite{gabi:cell}. In this work we utilize the
new functional formulation of the electronic structure problem
and extend it to relativistic case.

The
 basic idea
is to introduce another relevant variable in addition to the
density $\rho $ and the magnetic moment density $\bf m$, namely
the local Green function. The latter is defined by projecting the
full Green function onto a separate subset of correlated
``heavy'' orbitals distinguished by the orbital index $a$  and
the spin index $\sigma$ from a complete set   of orbitals $\chi
_{a}^\sigma({\bf r}-{\bf R})\equiv \chi _{a R}^\sigma$ of a
tight-binding representation which we assume for simplicity to be
orthogonal. The local Green function is therefore given by a
matrix $\hat{G}$ with elements \cite{chitra}
\begin{eqnarray} &&{\ }G_{ab}^{\sigma\sigma^\prime}(i\omega
,{\bf R})=-\left\langle c_{aR}^\sigma(i\omega
)c_{bR}^{\sigma^\prime +}(i\omega )\right\rangle = \label{Gab} \\
&&-\int \chi _{a}^{\sigma\ast }({\bf r-R})\left\langle \psi ({\bf
r},i\omega
)\psi ^{+}({\bf r}^{\prime },i\omega )\right\rangle \chi _{b}^{\sigma^\prime}({\bf r}%
^{\prime }-{\bf R})d{\bf r}d{\bf r}^{\prime }.  \nonumber
\end{eqnarray}
We then construct a functional $\Gamma \lbrack \,\rho ,{\bf
m},\hat{G}\,]$ which gives the exact free energy at a stationary
point.

To describe the new method,  we consider fermion system under an
external potential $V_{\mbox{\scriptsize ext}} $ and an external
magnetic field ${\bf h} $. For relativistic effects spin-orbit
coupling,  whose effects are important for magnetic anisotropy
calculations, are also considered. Spin-orbit coupling is
included according to the suggestion by Andersen~\cite{OA75}. It
is useful to introduce the notion of the Kohn-Sham potential
$V_{\mbox{\scriptsize KS}}  $, the Kohn-Sham magnetic field $ {\bf
h}_{\mbox{\scriptsize KS}} $ and its dynamical analog
$\Sigma^{\sigma\sigma'}_{ab}(i\omega_n)$. They are defined as the
functions that one needs to add to the kinetic energy matrix so
as to obtain a given density and spectral function of the heavy
orbitals namely: \begin{eqnarray} \rho(\R)=&&T\sum_{i\omega
_{n}}{\rm Tr}_s\left\langle \R s \left|[(i\omega _{n}+\nabla
^{2}/2-V_{\mbox{\scriptsize
KS}}){\mbox I} \right.\right. \nonumber \\
&&-\left.\left.2\mu_B{\bf s}\cdot {\bf h}_{\mbox{\scriptsize
KS}}-\xi({\bf r}){\bf l}\cdot{\bf s}-\Sigma ]^{-1}\right| \R s
\right\rangle e^{i\omega _{n}0^{+}}, \label{rholda+u}\\ {\bf
m}({\bf r})=&&-2\mu_BT\sum_{i\omega _{n}}{\rm Tr}_s\left\langle
{{\bf r}} s\left| {\bf s}[(i\omega _{n}+\nabla
^{2}/2-V_{\mbox{\scriptsize KS}}){\mbox I}
\right.\right.\nonumber\\&&\left.\left. -2\mu_B{\bf s}\cdot {\bf
h}_{\mbox{\scriptsize KS}}-\xi({\bf r}){\bf l}\cdot{\bf s}
-\Sigma]^{-1} \right| {\bf r} s \right\rangle e^{i\omega
_{n}0^{+}}, \label{mholda+u}
\end{eqnarray} where ${\rm Tr}_s$ is the trace over spin
space, ${\bf l}$ and ${\bf s}$ are one-electron orbital and spin
angular momentum operator, respectively. The spin angular
momentum operator is expressed in terms of Pauli matrices ${\bf
s}=\vec\sigma/2$ and  ${\mbox I}$ is $2\times 2$ unit matrix.
 $V_{KS}$ and ${\bf h}_{\mbox{\scriptsize KS}}$ are functions of
${\bf r}$. The chemical potential $\mu $ is set to zero throughout
the current section, and $\Sigma $ is given by
\begin{equation} \Sigma \equiv \Sigma ({\bf r},{\bf
r}^{\prime },i\omega )=\sum_{ab\sigma\sigma^\prime R}\chi
_{a}^{\sigma\ast }({\bf r-R})\Sigma _{ab}^{\sigma\sigma'}(i\omega )\chi _{ b}^{\sigma^\prime}({\bf r}^{\prime }{\bf %
-R}).  \nonumber \end{equation} $\xi (\R)$ determines the
strength of spin--orbit coupling and in practice is
determined~\cite{Harmon} by radial derivative of the $l=0$
component of the Kohn--Sham potential inside an atomic sphere:
\begin{equation} \xi
(r)=\frac{2}{c^{2}}\frac{dV_{\mbox{\scriptsize KS}}(r)}{dr}.
\label{sor} \end{equation}

When spin--orbit coupling is present, the intra--atomic
magnetization ${\bf m}(\R)$ is not collinear, i.e., the direction
of magnetization depends on the position $\R$. Therefore, the
magnetization must be treated as a general vector field, which
realizes non-collinear intra-atomic nature of this quantity. Such
general magnetization scheme has been recently discussed
\cite{singh}

In terms of these quantities  and the matrix of local
interactions $\hat{U}$, we write down the DMFT+LSDA functional:
\begin{eqnarray} &&\Gamma _{\mbox{\scriptsize{LSDA+DMFT}}}(\rho
,V_{\mbox{\scriptsize{KS}}},{\bf m}, {\bf h}_{\mbox{\scriptsize
KS}},\hat{G},\hat{\Sigma})= \nonumber
\\ &&-T\sum_{\omega }e^{i\omega 0^{+}}{\rm Tr}\log
[(i\omega +\nabla ^{2}-V_{\mbox{\scriptsize
KS}}){\mbox{I}}-2\mu_B{\bf s}\cdot{\bf h}_{\mbox{\scriptsize
KS}}\nonumber \\&& -\xi({\bf r}){\bf l}\cdot{\bf s}-\Sigma ]
-\int d{\bf r} V_{KS}({\bf
r})\rho ({\bf r})
+\int d\R  {\bf m}({\bf r})\cdot {\bf h}_{\mbox{\scriptsize
KS}}({\bf r}) \nonumber \\
&&-\sum_{\omega }e^{i\omega 0^{+}}%
{\rm Tr}[\hat{\Sigma}(i\omega )\hat{G}(i\omega )]+\int d{\bf r}
V_{ext}({\bf r})\rho ({\bf r})\nonumber \\ &&-\int
d\R {\bf h}({\bf r})\cdot {\bf m}({\bf r})+\frac{1}{2}\int d{\bf r}d{\bf r }^{\prime }  \frac{\rho (%
{\bf r})\rho ({\bf r}^{\prime })}{|{\bf r}-{\bf r}^{\prime }|}
\nonumber
\\&&+E_{\mbox{\scriptsize xc}}^{\mbox{\scriptsize
LSDA}}[\rho,{\bf m} ] +\sum_{R}[\Phi \lbrack \hat{G}]-\Phi _{DC}].
\label{functional} \end{eqnarray} $\Phi \lbrack \hat{G}]$ is the
sum of the two--particle irreducible local diagrams constructed
with the local interaction matrix $\hat{U}$, and the local heavy
propagator $\hat{G}$. $\Phi _{DC}$ is the so--called double
counting term which subtracts the average energy of the heavy
level already described by LDA. Expression (\ref{functional})
ensures that the Greens function obtained from its extremization
will satisfy the Luttinger theorem.

$E_{\mbox{\scriptsize xc}}^{\mbox{\scriptsize LSDA}}[\rho,{\bf
m}]$ is the LSDA exchange correlation energy. Since the exact
exchange correlation energy functional
  is not known, the usefulness of this
approach is due to the existence of successful approximations to
the exchange correlation energy functional as Kohn and Sham
proposed.   When nontrivial magnetic moment is present, the
exchange correlation energy functional is assumed to be dependent
on density and magnetization: \begin{eqnarray}
E_{\mbox{\scriptsize xc}}^{\mbox{\scriptsize LSDA}}[\rho,{\bf
m}]&=&\int d\R \epsilon_{\mbox{\scriptsize xc}}[\rho(\R), { m}(\R)
]\rho(\R)\\ &+&\int d\R f_{\mbox{\scriptsize xc}}[\rho(\R), m(\R)
] { m}(\R)(\R) ,\label{lsda-o} \end{eqnarray} where $m=|{\bf m}|$.

The functional (\ref{functional}) can be viewed as a functional
of six independent variables, since the stationary condition in
the conjugate fields reproduces the definition of the dynamical
potential and the Weiss field. Extremizing it with respect
$V_{\mbox{\scriptsize KS}}$, ${\bf h}_{\mbox{\scriptsize KS}}$,
and $\Sigma$ lead us to compute the density(Eq.\
(\ref{rholda+u})), the magnetic moment density(Eq.\
(\ref{mholda+u})), and the Green function
$G_{ab}^{\sigma\sigma^\prime}(i\omega )$ (Eq.\ (\ref{Gab2})),
respectively. The Kohn--Sham potential $V_{\mbox{\scriptsize
KS}}(\R)$ and Kohn--Sham magnetic field ${\bf
h}_{\mbox{\scriptsize KS}}({\bf r})$ are obtained by extremizing
the functional with respect to $\rho(\R)$ and ${\bf m}(\R)$:
\begin{eqnarray} V_{\mbox{\scriptsize KS}}({\bf
r})&=&V_{\mbox{\scriptsize ext}}({\bf r})+\int
d\R'\frac{\rho(\R')}{\left|{\bf r}-\R'\right|} +\frac{\delta
E_{\mbox{\scriptsize xc}}^{\mbox{\scriptsize LDA}} [\rho,{\bf
m}]}{\delta \rho({\bf r})}, \label{vks} \\ {\bf
h}_{\mbox{\scriptsize KS}}({\bf r})&=&{\bf h}({\bf r})
+\frac{\delta E_{\mbox{\scriptsize xc}}^{\mbox{\scriptsize
LDA}}[\rho,{\bf m}]}{\delta {\bf m}({\bf r})}. \label{hks}
\end{eqnarray} Extremizing with respect to
$G_{ab}^{\sigma\sigma^\prime}$ yields the equation for self
energy \begin{equation} \Sigma _{ab}^{\sigma\sigma^\prime}
(i\omega )=\frac{\delta \Phi }{\delta G_{ab}^{\sigma\sigma^\prime}(i\omega )}-\frac {%
\delta \Phi _{DC}}{\delta G_{ab}^{\sigma\sigma^\prime}(i\omega )}.
\label{sigmaabw}
\end{equation}

The physical meaning of the dynamical potential $\Sigma $ is
parallel to the meaning of the original Kohn-Sham potential
$V_{\mbox{\scriptsize KS}}$: it is the function that one needs to
add to the correlated block of the one-electron Hamiltonian in
order to obtain the desired local Green function: \begin{equation}
G_{ab}^{\sigma\sigma^\prime}(i\omega )=\sum_{{\bf k}}\,[i\omega -\hat{H}^{{\bf k}}-\hat{\Sigma}%
(i\omega )]_{ab}^{\sigma\sigma^\prime-1},  \label{Gab2}
\end{equation} where $H_{ab}^{\sigma\sigma^\prime{\bf
k}}=\langle \chi _{a{\bf k}}^\sigma|(-\nabla
^{2}+V_{\mbox{\scriptsize KS}}){\mbox I}
+2\mu_B{\bf s}\cdot{\bf h}_{\mbox{\scriptsize KS}}+\xi({\bf r}){\bf l}\cdot{\bf s}|\chi _{b%
{\bf k}}^{\sigma^\prime}\rangle $ is the one-electron Hamiltonian
in ${\bf k}$-space. It is the frequency dependence of the
dynamical potential which allows us to treat Hubbard bands and
quasiparticle bands on the same footing.

In general, an explicit form of $\Phi[G]$ is not available. DMFT
maps the DMFT$+$LSDA function to an Anderson impurity model.
Self--consistency equations obtained in this way are used to find
the self energy~(\ref{sigmaabw}). In this paper we confine
ourselves to zero temperature and make an additional assumption
on solving the impurity model using the Hartree--Fock
approximation. In this limit an explicit form of $\Phi[G]$ is
available and DMFT self--consistency loop is unnecessary. We
first figure out the Coulomb interaction   by considering a
Hartree--Fock averaging of the original expression for the
Coulomb interaction
  given by \begin{equation}  \frac{1}{2}\sum_{\sigma
\sigma ^{\prime }}\sum_{abcd}\langle a\sigma b\sigma ^{\prime
}|\frac{e^{2}}{r}|c\sigma d\sigma^{\prime }\rangle c_{a}^{\sigma
+}c_{b}^{\sigma^{\prime }+}c_{d}^{\sigma^\prime}c_{c}^{\sigma }.
\label{coulomb}
\end{equation}
In this limit, the sum of local graphs reduce to
\begin{eqnarray}
\Phi[\hat G]&=&\frac{1}{2}\sum_{abcd\sigma
}U_{abcd}n_{ab}^{\sigma \sigma }n_{cd}^{-\sigma -\sigma }
\nonumber
\\ &+&\frac{1}{2}%
\sum_{abcd\sigma }(U_{abcd}-J_{abcd})n_{ab}^{\sigma \sigma
}n_{cd}^{\sigma \sigma } \nonumber \\
&-&\frac{1}{2}\sum_{abcd\sigma }J_{abcd}n_{ab}^{\sigma -\sigma
}n_{cd}^{-\sigma \sigma }. \label{Emodel}
\end{eqnarray} Here, the matrices $U_{abcd}=\langle
ac|v_C|bd\rangle$ and $J_{abcd} =\langle ac|v_C|db\rangle$ have
the following definitions: \begin{eqnarray} U_{abcd} &=& \int\chi
_{a}^{\sigma\ast }(\R)\chi _{c}^{\sigma\ast }(\R')v_{C}(\R-{\bf
r}')\chi _{b}^\sigma(\R)\chi _{d}^\sigma(\R')d\R d\R',\nonumber
\\ J_{abcd} &=& \int \chi_a (\R)^{\sigma\ast }\chi
_{c}^{\sigma\ast }(\R')v_{C}(\R-\R')\chi _{d}^\sigma(\R)\chi
_{b}^\sigma(\R')d\R d\R',\nonumber \end{eqnarray} where the
Coulomb interaction $v_{C}(\R-\R')$ has to take into account the
effects of screening by conduction electrons. Note that the
matrices $U_{abcd}$ and $J_{abcd}$ are spin independent since the
Coulomb interaction is independent of spin. The occupancy matrix
$n_{ab}^{\sigma \sigma ^{\prime }}$ is a derived quantity of the
Green function:
\begin{equation} n_{ab}^{\sigma \sigma ^{\prime
}}=T\sum_{\omega}G_{ab}^{\sigma\sigma^\prime}(i\omega )e^{i\omega
0^{+}}. \label{occu} \end{equation} Notice that when spin--orbit
coupling is taken into account, the occupancy matrix becomes
non--diagonal with respect to spin index even though the
interaction matrices $U_{abcd}$ and $J_{abcd}$ are spin
independent.

The self energy $\Sigma_{ab}^{\sigma\sigma^\prime}$ now takes the
from for spin diagonal elements
\begin{eqnarray}
\Sigma _{ab}^{\sigma \sigma }&=&\sum_{cd}U_{abcd}n_{cd}^{-\sigma
-\sigma }+\sum_{cd}(U_{abcd}-J_{abcd})n_{cd}^{\sigma \sigma }
\nonumber \\ &-&\frac {%
\delta \Phi _{DC}}{\delta G_{ab}^{\sigma\sigma}(i\omega
)},\label{lambdasigma} \end{eqnarray} and for spin off--diagonal
elements it is given by \begin{equation} \Sigma _{ab}^{\sigma
-\sigma }=-\sum_{cd}J_{abcd}n_{cd}^{-\sigma
\sigma } -\frac {%
\delta \Phi _{DC}}{\delta G_{ab}^{\sigma-\sigma}(i\omega )}.
\label{lambdaoff} \end{equation} The off--diagonal elements of
the self energy only present when spin--orbit coupling is
included, hence a relativistic effect. To make it more physically
transparent we can introduce magnetic moments at the given shell
by \begin{equation} m_{ab}^{\mu }=\sum_{\sigma \sigma ^{\prime
}}s_{\sigma \sigma ^{\prime }}^{\mu }n_{ab}^{\sigma \sigma
^{\prime }} \label{magmom}
\end{equation}
where $\mu $ runs over $x,y,z$ for Cartesian coordinates, or over,$-1,0,+1$ (%
$z,\pm $) for spherical coordinates. Relativistic correction from
strong correlations can be written in physically transparent form
\begin{eqnarray}
\frac{1}{2}\sum_{abcd\sigma }J_{abcd}n_{ab}^{\sigma -\sigma
}n_{cd}^{-\sigma \sigma }&\equiv&
\frac{1}{2}\sum_{abcd}m_{ab}^{(+)}J_{abcd}m_{cd}^{(-)} \nonumber
\\
&+&\frac{1%
}{2}\sum_{abcd}m_{ab}^{(-)}J_{abcd}m_{cd}^{(+)}  \label{phystrans}
\end{eqnarray}
and in principle has room for further generalization of exchange
matrix $J_{abcd}$ to be anisotropic, i.e depend on $\mu \mu
^{\prime }$: $J_{abcd}^{\mu \mu ^{\prime }}.$

Part of the energy added by $\Phi[\hat G]$ is already included in
LSDA functional. The double counting term
$\Phi_{\mbox{\scriptsize dc}}$ is added to subtract this already
included part of $\Phi[\hat G]$.   It was proposed~\cite{afl}
that the form for $\Phi[\hat G]$ is
\begin{equation} \Phi_{\mbox{\scriptsize
dc}}^{\mbox{\scriptsize Model}}
=\frac{1}{2}\bar{U}\bar{n}(\bar{n}-1)-\frac{1}{2}\bar{J}[\bar{n%
}^{\uparrow }(\bar{n}^{\uparrow }-1)+\bar{n}^{\downarrow }(\bar{n}%
^{\downarrow }-1)],  \label{C6} \end{equation} where
\begin{eqnarray} \bar{U}
&=&\frac{1}{(2l+1)^{2}}\sum_{ab}\langle
ab|\frac{1}{r}|ab\rangle, \label{C7} \\
\bar{J} &=&\bar{U}-\frac{1}{2l(2l+1)}\sum_{ab}(\langle ab|\frac{1}{r}%
|ab\rangle -\langle ab|\frac{1}{r}|ba\rangle ),  \label{C8}
\end{eqnarray} and   $\bar{n}^{\sigma }=\sum_{a}n_{aa}^{\sigma },$ and $\bar{n}=\bar{n}%
^{\uparrow }+\bar{n}^{\downarrow }$. The subtraction by $1$ is
made to take the self-interaction into account. This generates
the self energy in the form: \begin{eqnarray} \Sigma
_{ab}^{\sigma \sigma }&=&\sum_{cd}U_{abcd}n_{cd}^{-\sigma -\sigma
}+\sum_{cd}(U_{abcd}-J_{abcd})n_{cd}^{\sigma \sigma }
\nonumber \\&-&\delta_{ab}\bar{U}(\bar{n}-\frac{1}{2}%
)+\delta_{ab}\bar J(\bar{n}^{\sigma }-\frac{1}{2}) \label{potldau}
\\ \Sigma _{ab}^{\sigma -\sigma }&=&-\sum_{cd}J_{abcd}n_{cd}^{-\sigma
\sigma }.
\end{eqnarray}

As an example, when only the effect of $U$ is under
investigation, the $U$ and $J$ matrices are  $U_{abcd}=
\delta_{ab}\delta_{cd}U $, $J_{abcd}= \delta_{ad}\delta_{cb}U$,
$\bar U=U$, and $\bar J=0$. This simple $U$ and $J$ matrices make
it possible to write down corrections to LSDA functional and LSDA
Kohn-Sham potential: \begin{eqnarray} \Phi[\hat
G]-\Phi_{\mbox{\scriptsize dc}}^{\mbox{\scriptsize
Model}}&=&-\frac{1}{2}\sum_{\sigma}\sum_{ab}U(n_{ab}^{\sigma\sigma
}n_{ba}^{\sigma\sigma }+n_{ab}^{\sigma-\sigma
}n_{ba}^{-\sigma\sigma })\nonumber\\ &&-\frac{1}{2} U \bar n
\end{eqnarray}
\begin{eqnarray} \Sigma _{ab}^{\sigma\sigma } &=&
U(\frac{1}{2}\delta_{ab}-n_{ba}^{\sigma\sigma}) \\
\Sigma_{ab}^{\sigma-\sigma } &=& Un_{ba}^{-\sigma\sigma}
\end{eqnarray}

The DMFT self consistency equation identifies the Green function
of the original model and the Green function of the mapped
impurity model to find the self energy. Now that we can express
the sum of local graphs $\Phi[\hat G]$ in terms of the original
Green function, the DMFT loop need not to be performed. The
problem is now reduced to extremizing  the functional
[Eq.~(\ref{functional})] with the expression for the sum of local
graphs [Eq.~(\ref{Emodel})], which is exactly the LDA$+$U
method~\cite{anisimov}.

The DMFT functional and its static correspondent LDA$+$U
functional are defined once a set of projectors
$\{\chi_a^\sigma(\R)\}$ and a matrix of interactions $U_{abcd}$
and $J_{abcd}$ are prescribed.   When $l$ orbitals are used as
the projection operators, the matrix is expressed in terms of
Slater parameters $F^k$. For $a\equiv lm,b\equiv lk,c\equiv
l^{\prime }m^{\prime },d\equiv l^{\prime }k^{\prime }$ and
representing $\chi_{a}^\uparrow({\bf r}) =\phi _{lm}({\bf
r})(1,0)^{\mbox{\scriptsize T}}$, where $\phi _{lm}({\bf r})=\phi
_{l}(r)i^{l}Y_{lm}(\hat{r}),$ we can express the matrices
$U_{abcd}$ and $ J_{abcd}$ in the following manner:
\begin{eqnarray} \langle lml^{\prime }m^{\prime
}|\frac{1}{r}|lkl^{\prime }k^{\prime }\rangle
&=&\sum_{l^{\prime \prime }=0,2,...}^{\min (2l,2l^{\prime })}\frac{4\pi }{%
2l^{\prime \prime }+1}F_{ll^{\prime }}^{(u)l^{\prime \prime }}
\\&\times& (-1)^{m''}C_{lklm}^{l^{\prime \prime }m^{\prime \prime
}=m-k}C_{l^{\prime }m^{\prime }l^{\prime }k^{\prime }}^{l^{\prime
\prime }m^{\prime \prime }=k^{\prime
}-m^{\prime }}  \nonumber \\
\langle lml^{\prime }m^{\prime }|\frac{1}{r}|l^{\prime }k^{\prime
}lk\rangle
&=&\sum_{l^{\prime \prime }=0,2,...}^{\min (2l,2l^{\prime })}\frac{4\pi }{%
2l^{\prime \prime }+1}F_{ll^{\prime }}^{(j)l^{\prime \prime }} \\
&\times& (-1)^{m''}C_{l^{\prime }k^{\prime }lm}^{l^{\prime \prime
}m^{\prime \prime }=m-k^{\prime }}C_{l^{\prime }m^{\prime
}lk}^{l^{\prime \prime }m^{\prime \prime }=k-m^{\prime }}\nonumber
\end{eqnarray}    where the quantities
$C_{LL^{\prime }}^{L^{\prime \prime }}$ are the Gaunt
coefficients which are the integrals of the products of three
spherical harmonics \begin{equation} C_{LL^{\prime }}^{L^{\prime
\prime }}=\int Y_{L}({\bf \hat{r}})Y_{L^{\prime }}^{\ast }({\bf
\hat{r}})Y_{L^{\prime \prime }}({\bf \hat{r}})d{\bf \hat{r}}.
\end{equation} The quantities $F^{(u)}$ and $F^{(j)}$ are given by the
following radial integrals \begin{eqnarray}
F_{ll^{\prime }}^{(u)l^{\prime \prime }} &=&\int \frac{r^{l^{\prime \prime }}%
}{r^{\prime l^{\prime \prime }+1}}\phi _{l}^{2}(r)\phi
_{l^{\prime }}^{2}(r^{\prime })drdr^{\prime } \\
F_{ll^{\prime }}^{(j)l^{\prime \prime }} &=&\int \frac{r^{l^{\prime \prime }}%
}{r^{\prime l^{\prime \prime }+1}}\phi _{l}(r)\phi _{l^{\prime
}}(r)\phi _{l}(r^{\prime })\phi _{l^{\prime }}(r^{\prime
})drdr^{\prime }. \end{eqnarray} When $l\equiv l^{\prime }$, the
quantites $F^{(u)}$ and $F^{(j)}$ are equal and have a name of
Slater integrals which for s--electrons are reduced to one
constant $F^{(0)},$ for p--electrons there are two constants:
$F^{(0)},$ $F^{(2)},$ for d's: $F^{(0)},$ $F^{(2)},$ $F^{(4)},$
etc. In this case, the expressions for $U$ and $J$ are reduced to
\begin{equation} \langle m, m''|v_C|m',m'''\rangle
=\sum_ka_k(m,m',m'',m''')F^k, \end{equation} where $0\le k \le
2l$, and \begin{eqnarray}  &&a_k(m,m',m'',m''')= \nonumber \\ &&
\frac{4\pi}{2k+1}(-1)^{q}C_{lmlm'}^{kq=m-m'}
C_{lm''lm'''}^{kq=m'''-m''} .
\end{eqnarray} Slater integrals can be linked to Coulomb
and Stoner parameters $U$ and $J$ obtained from LSDA supercell
procedures via $U=F^0$ and $J=(F^2+F^4)/14$. The ratio $F^2/F^4$
is to a good accuracy a constant $\sim 0.625$ for $d$ electrons.
For $f$ electrons, the corresponding expression is $U=F^0$ and
$J=(286F^2+195F^4+250F^6)/6435$.

To summarize, we have shown the equivalence of Hartree--Fock
approximation of DMFT and LDA$+$U method. LDA$+$U is the method
proposed to overcome the difficulties of LDA when strong
correlations are present~\cite{aza}. Since the density uniquely
defines the Kohn--Sham orbitals, and they in turn, determine the
occupancy matrix of the correlated orbitals, once a choice of
correlated orbitals is made, we still have a functional of the
density alone. However it is useful to proceed with Eq.
(\ref{functional}), and think of the LDA + U functional as a
functional of $\rho$, $V_{\mbox{\scriptsize KS}}$, ${\bf m}$,
${\bf h}_{\mbox{\scriptsize KS}}$, $G^{\sigma\sigma' }_{ab}$, and
$\Sigma^{\sigma\sigma' }_{ab}$, whose minimum gives better
approximations to the ground--state energy in strongly correlated
situations. Allowing the functional to depend on the projection
of the Kohn--Sham energies onto a given orbital, allows the
possibility of orbitally ordered states. \ This is a major
advance over LDA in situations where this orbital order is
present. As recognized many years ago, this is a very efficient
way of gaining energy in correlated situations, and is realized
in a wide variety of systems.

\section{Numerical Calculation of MAE} \label{maesec}

We calculate MAE  by taking the difference of two total energies
with different directions of magnetization [MAE=$E(111)-E(001)$].
The total energies are obtained via fully self consistent
solutions. Since the total energy calculation requires high
precision, full potential LMTO method~\cite{Sav} has been
employed. For the $\vec{k}$ space integration, we follow the
analysis given by Trygg and co--workers~\cite{tjew} and use the
special point method~\cite{froyen} with a Gaussian broadening~\cite{mp} of $%
15\ mRy$. The validity and convergence of this procedure has been
tested in their work ~\cite{tjew}. For convergence of the total
energies within
desired accuracy, about $15000\ k$-points per Brillouine zone are needed. We used $28000\ k$%
-points to reduce possible numerical noise, while the convergency
is tested up to $84000k$-points. Our calculations include
non-spherical terms of the charge density and potential both
within the atomic spheres and in the interstitial
region~\cite{Sav}. All the low-lying semi-core states are treated
together with the valence states in a common Hamiltonian matrix in
order to avoid unnecessary uncertainties. These calculations are
spin polarized and assume the existence of long-range magnetic
order. Spin-orbit coupling is implemented according to the
suggestions by Andersen~\cite{OA75}. We also treat magnetization
as a general vector field, which realizes non-collinear
intra-atomic nature of this quantity. Such general magnetization
scheme has been recently discussed\cite{singh}.

  We first test our method in case of LDA ($%
U=J=0$). To compare with previous calculations, we turn off the
non-collinearity of magnetization which makes it collinear with
the quantization axis. The calculation gives correct orders of
magnitude for both fcc Ni ($0.5\ \mu eV$) and bcc Fe ($0.5\ \mu
eV$) but with the wrong easy axis for Ni, which is the same
result as the previous one~\cite{tjew}. Turning on the
non-collinearity results in a a larger value of the absolute
value of the MAE ($2.9\ \mu eV$) for Ni but the easy axis
predicted to be $(001)$ which is still wrong. The magnitude of
the experimental MAE of Ni is $2.8\ \mu eV$ and the easy axis is
aligned along $(111)$ direction~\cite{landolt}. For Fe, the
non-collinearity of magnetization changes neither MAE ($0.5\ \mu
eV$) nor the easy axis $(001)$ from the collinear result. The
magnitude of the experimental MAE of Fe is $1.4\ \mu eV$ and the
easy axis is aligned along $(001)$ direction.

\begin{figure}[htb]
\epsfig{file=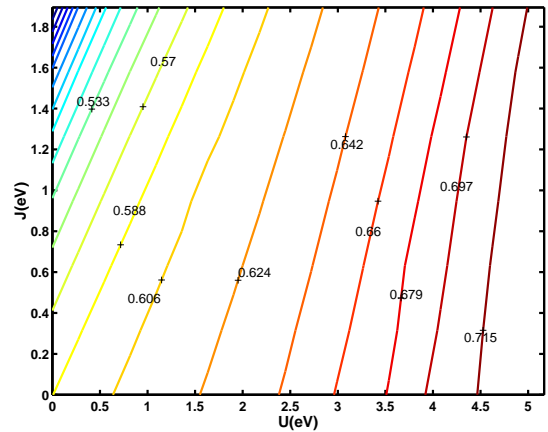,width=0.4\textwidth}
\vspace{6pt}
\caption{Ni. Contour plot of magnetic moment as a function
of   $U$ and   $J$.
The
contour is drawn at $0.018\ \mu_{\mbox{\scriptsize B}}$
interval, which is $2.9\%$ of the experimental value of
magnetic moment $0.606\ \mu_{\mbox{\scriptsize B}}$.
}
\label{fig:magNiOUJ} 
\end{figure}

\begin{figure}[thb]
\epsfig{file=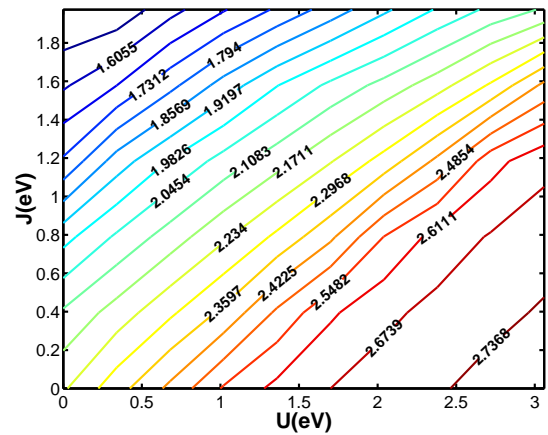,width=0.4\textwidth}
\vspace{6pt}
\caption{Fe. Contour plot of magnetic moment as a function
of   $U$ and   $J$.
The
contour is drawn at $0.063\ \mu_{\mbox{\scriptsize B}}$
interval, which is $2.9\%$ of the experimental value of
magnetic moment $2.2\ \mu_{\mbox{\scriptsize B}}$.
}
\label{fig:magFeOUJ} \end{figure}

We now describe the effect of correlations, which is crucial in
predicting the correct axis of Ni. We first study the effects of
strong correlations on magnetic moments. We have scanned the $(U,
J)$ parameter space to obtain magnetic moment as a function of
$U$ and $J$ (see Figs.\ \ref{fig:magNiOUJ} and
\ref{fig:magFeOUJ}). Magnetic moment increases as $U$ increases,
but decreases as $J$ increases for both Ni and Fe. The magnetic
moments of Ni and Fe change up to $20\%$ in the parameter range.
Notice that well--defined contours exist where the magnetic
moments are close to experimental values for both Ni and Fe. In
comparison, the magnetic moment of Fe depends more strongly on
$J$ than that of Ni. This result is in agreement with an earlier
work \cite{Kudrnovsky}.

We now discuss our calculated MAE. The load of computing MAE is
very heavy. Rather than calculating the quantities in the $(U,J)$
parameter space, we follow three paths: two paths with constant
$J$, and  the path with experimental magnetic moments.

\begin{figure}[htb]
\epsfig{file=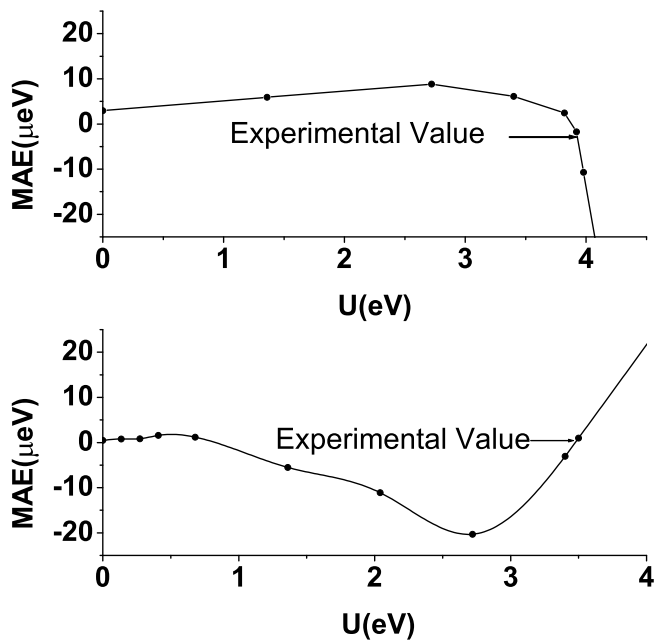, width=0.48\textwidth}
\vspace{6pt}
\caption{The
magneto--crystalline anisotropy energy $\mbox{MAE}=E(111)-E(001)$
for Ni (top) and Fe (bottom) as
functions of $U$. The experimental MAEs are marked by arrows for
Fe ($1.4\ \protect\mu eV$) and Ni ($-2.8\ \protect\mu eV$).   The
value of $J$ is held to $0$.
}
\label{mae} 
\end{figure}

We first walk along a path with $J=0\ eV$. For Ni (see Fig.\
\ref{mae}), as U increases, the MAE of Ni smoothly increases
until $U$ reaches $2.5\ eV$ and then smoothly decreases up to the
value $3.8\ \mu eV$. Around $U=3.9\ eV$, the MAE decreases
abruptly to negative value. Around $U=4.0\ eV$, the experimental
order of magnitude and the correct easy axis (111) are restored.
The change from the wrong easy axis to the correct easy axis
occurs over the range of $\delta U\sim 0.2 eV$, which is the
order of spin-orbit coupling constant ($\sim 0.1 eV$).

For  Fe, the MAE decreases on increasing $U$ to negative values,
where the magnetization takes the wrong axis. From $U=2.7\ eV$,
it increases back to the correct direction of easy axis (positive
MAE). Around $U=3.5\ eV$, it restores the correct easy axis and
the experimental value of MAE is reproduced.

In this study on effects of intra--atomic repulsion, we see that
it is possible to predict the correct magnetic anisotropy at a
nontrivial $U$. Notice, however, that the calculated magnetic
moments are about $20\%$ larger than those of experiments for
both Ni and Fe. As seen in the Figs.\ \ref{fig:magNiOUJ} and
\ref{fig:magFeOUJ}, the magnetic moment decreases as $J$
increases. This lead us to follow a path with  nontrivial $J$.

\begin{figure}[htb]
\epsfig{file=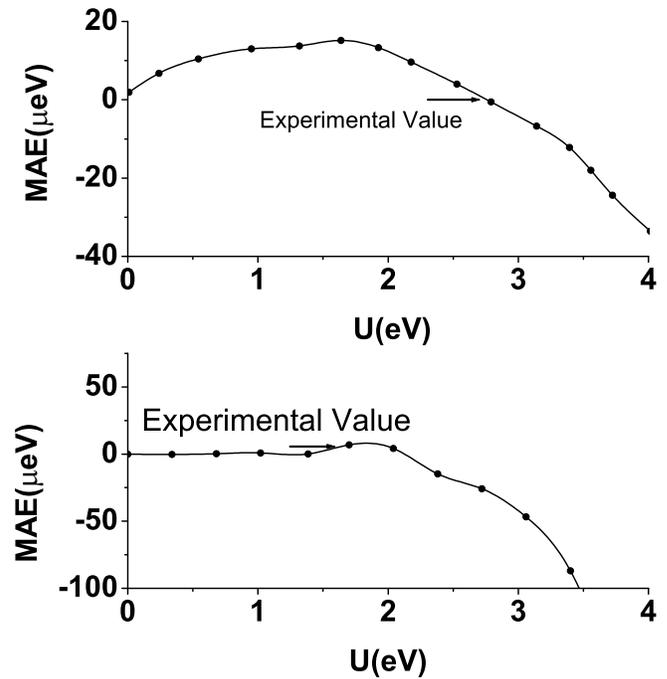, width=0.48\textwidth}
\vspace{6pt}
\caption{The
magneto--crystalline anisotropy energy $\mbox{MAE}=E(111)-E(001)$
 for Ni (top) and Fe (bottom) as
functions of $U$. The experimental MAEs are marked by arrows for
Fe ($1.4\ \protect\mu eV$) and Ni ($-2.8\ \protect\mu eV$).   The
values of $J$ are fixed to $0.9$ for both Ni and Fe.
}
\label{mae_no_j9} \end{figure}

We pick $J=0.9\ eV$ that seems to be a canonical value of $J$. For
Ni, (see Fig. \ref{mae_no_j9}), as $U$ increases, MAE increases
to $15\ \mu eV$ till $U=1.6\ eV$. Then MAE decreases, changing
the easy axis from $(001)$ to $(111)$ at $U=2.6\ eV$. The
experimental value of MAE with the correct easy axis is predicted
at $2.7\ eV$.

For Fe, MAE increases to $10\ \mu eV$ till $U=1.9\ eV$. Then MAE
decreases, changing the easy axis to $(111)$ direction at $U=2.2\
eV$. The experimental value of MAE with the correct easy axis is
predicted at $1.7\ eV$.

In this study of effects of intra--atomic repulsion ($U$) with
nontrivial $J$, we are also able to predict the correct magnetic
anisotropy at a nontrivial $U$, much less than that of the case
with trivial $J$. Notice that the calculated magnetic moments are
about $5\%$ larger than those of experiments, a big improvement
over the case with $J=0$.

To treat properly the correlation effects on calculated
anisotropy energy, the intra--atomic repulsion $U$ and exchange
$J$ should be taken into account spontaneously. We, therefore,
follow a path in $(U, J)$ space determined by fixed values of
magnetic moments close to experimental value.

\begin{figure}[htb]
\epsfig{file=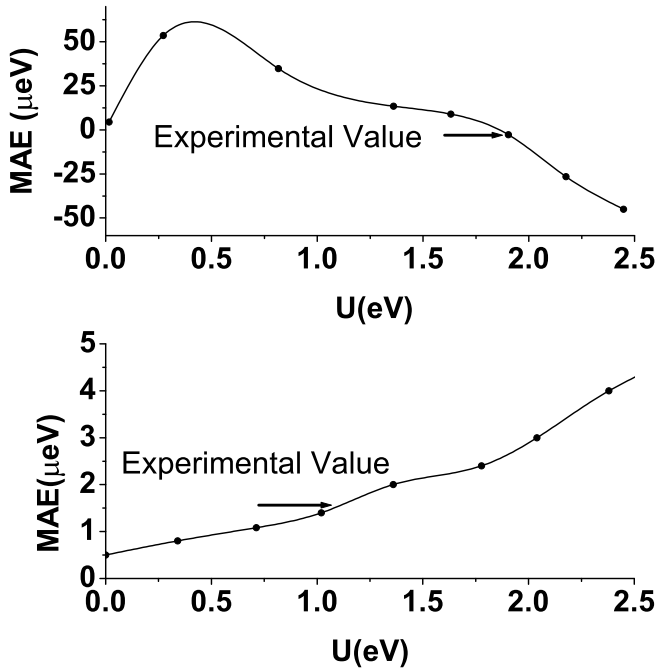,width=0.48\textwidth}
\caption{The
magneto--crystalline anisotropy energy $\mbox{MAE}=E(111)-E(001)$
functions of $U$. The experimental MAEs are marked by arrows for
Fe ($1.4\ \protect\mu eV$) and Ni ($-2.8\ \protect\mu eV$).   The
values of exchange parameter $J$ for every value of $U$ are
chosen to hold the magnetic moment of $0.61\ \protect\mu _B$ in
Ni and 2.2 $\protect\mu _B$ in Fe} 
\label{maeo3} 
\end{figure}

For Ni, we walked along the path of parameters $U$ and $J$ which
hold the magnetic moment to $0.61\ \mu _{B}$. The MAE first
increases to $60\ \mu eV$ ($U=0.5\ eV$, $J=0.3\ eV$) and then
decreases (see Fig.\ \ref{maeo3}). While decreasing it makes a
rather flat area from $U=1.4\ eV$, $J=0.9\ eV$ to $U=1.7\ eV$,
$J=1.1\ eV$ where MAE is positive and around $10\ \mu eV$. After
the flat area, the MAE changes from the wrong easy axis to the
correct easy axis. The correct magnetic anisotropy is predicted
at $U =1.9\ eV$ and $J=1.2\ eV$. The change from the wrong easy
axis to the correct easy axis occurs over the range of $\delta
U\sim 0.2eV$, which is of the order of spin-orbit coupling
constant ($\sim 0.1eV$).

For Fe, the MAE is calculated along the path of $U$ and $J$
values where the magnetic moment is fixed to $2.2\ \mu _{B}$. At
$U=0\ eV$ and $J=0\ eV$, the MAE is $0.5\ \mu eV$. It increases
as we move along the contour in the direction of increasing $U$
and $J$. The correct MAE with the correct direction of magnetic
moment is predicted at $U=1.2\ eV$ and $J=0.8\ eV$.


In this study of effects of $U$ and $J$, it  is again possible to
predict the correct magnetic anisotropy. What is more is that the
magnetic moment comes out to be the experimental values at the
same time.

We have demonstrated that it is possible to perform highly
precise calculation of the total energy in order to obtain both
the correct easy axes and the magnitudes of MAE for Fe and Ni.
This has been accomplished by including the strong correlation
effects via taking intra--atomic repulsion and exchange into
account, and incorporating the non--collinear magnetization.

\begin{figure}[htb]
\epsfig{file=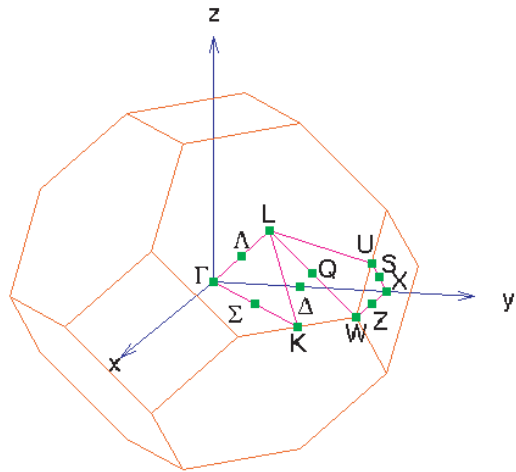, width=0.4\textwidth}
\vspace{6pt}
\caption{Brillouine Zone of fcc crystal structure.}
\label{bz}
\end{figure}

\begin{figure}[htb]
\epsfig{file=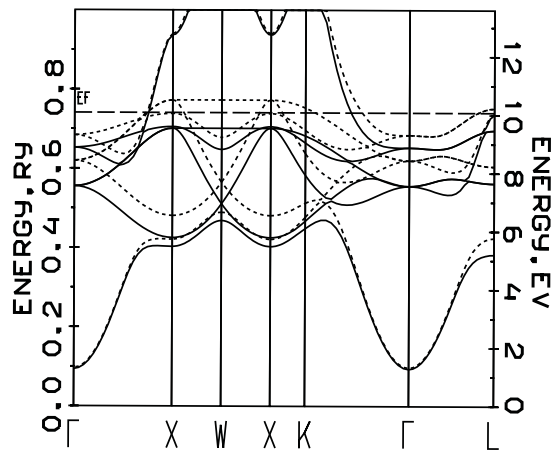, width=0.4\textwidth}
\vspace{6pt}
\caption{Calculated band structure of  Ni at $U=1.9\ eV$
and $J=1.2\ eV$.
The solid and dotted lines
correspond to majority and minority dominant spin carriers.
}
\label{band1}
\end{figure}

\begin{figure}[htb]
\epsfig{file=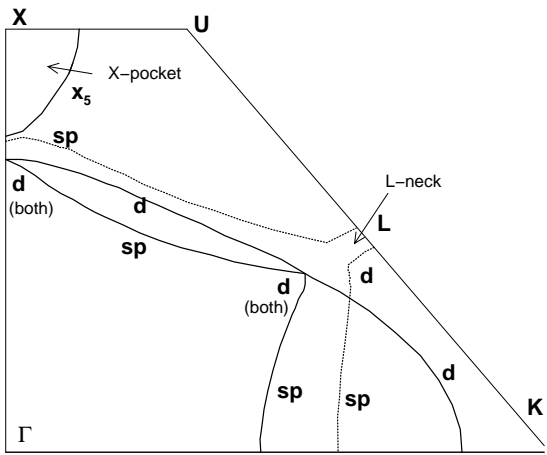, width=0.4\textwidth}
\vspace{6pt}
\caption{Calculated Fermi Surface of  Ni at $U=1.9\ eV$
and $J=1.2\ eV$.
The solid and dotted lines
correspond to majority and minority dominant spin carriers.
Dominant orbital characters are expressed.
Both experimentally confirmed $X_5$ pocket and $L$ neck can be seen.
The $X_2$ pocket is missing, which is in agreement with
experiments.
} 
\label{fermi1}
\end{figure}

\section{MAE and Band Structure}
\label{bandsec} We now present implications of our results on the
calculated electronic structure for   Ni. Fig.\ \ref{bz} shows
Brillouine zone and Fig.\ \ref{band1} shows band structure of Ni.
We now discuss the band structure at the $\Gamma$ point. From the
below we see three majority bands and three minority bands. The
lowest one is dominated by $4sp$ orbital, the middle one is
dominated by $t_{2g}$ orbital, and the highest one is dominated
by $e_g$ orbital. In this paragraph, we call the bands by its
dominant orbital character at the $\Gamma$ point. The $e_g$ band
has two subbands. One of the two subbands is strongly hybridized
with $sp$ bands from the middle of the $\Gamma X$ line. This band
can be easily identified since the energy grows rapidly from the
middle of the $\Gamma X$ line and the rapidly growing part is
dominated by $sp$ orbitals. Inspecting the $sp$ band, we observe
that the band is rather flat at later part of the $\Gamma X$ line.
In this area, the band is dominated by $e_g$ orbitals. The other
subband of $e_g$ band remains unhybridized along  the $\Gamma X$
line. The $t_{2g}$ band has two subbands in the $\Gamma X$
direction. One of the subbands is doubly degenerate and the other
band is singlet. The doubly degenerate band can be identified by
looking at the $X$ point since this band breaks into two subbands
there. Fig.\ \ref{fermi1} shows the Fermi surface deduced from
the band structure.

\begin{figure}[htb]
\centering
\epsfig{file=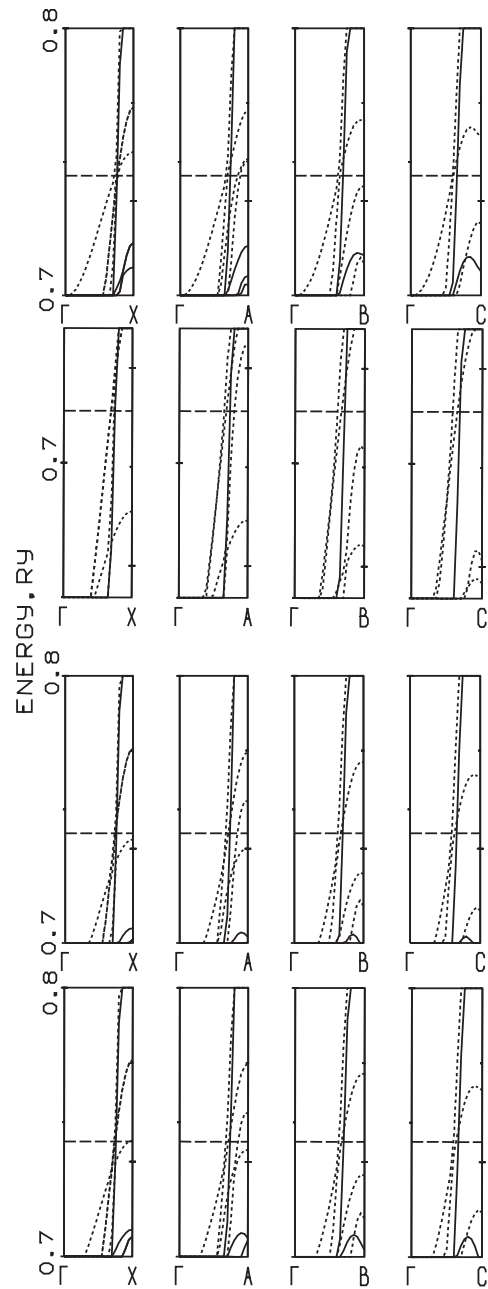,width=0.35\textwidth}
\caption{The X pockets. From the top, band structure near $X$ of LDA,
LDA$+$U at $U=4\ eV$ $J=0\ eV$, LDA$+$U at $U=1.9\ eV$ $J=1.2\ eV$, and
LDA$+$U at $U=1.7\ eV$ $J=1.1\ eV$. The size of the
windows are $1\ Ry$.
The solid lines represent bands with dominating up-spin.
The dotted lines represent bands with dominating
down-spin.
$X=(
 0,0,1)$,
$A= (1/16,1/16,1)$, $B=(
 2/16,2/16,1)$,
        $C
 =(3/16,3/16,1)$.
 } \label{FinX}
\end{figure}

We now discuss the band structure in relation with magnetic
anisotropy. One important feature which emerges from the
calculation is the absence of the $X_2$ pocket (see Fig.\
\ref{fermi1}). This has been predicted by LDA but has not been
found experimentally~\cite{wc}. Inspecting the four band
structures in band structure figures near the Fermi surface (Fig.\
\ref{FinX}), we notice that as we move away from the $\Gamma X$
direction to $\Gamma L$ direction, there is a band, which is
above the Fermi surface at the $X$ point, is submerged below the
Fermi level. The band is the doubly degenerate $t_{2g}$ subband
and  the pocket generated by this band is the experimentally
confirmed $X_5$ pocket.

In the LDA band structure (see the top of Fig.\ \ref{FinX}), there
is another band forming another $X$ pocket. This is the
unhybridized $e_g$ subband and the pocket generated by this band
is the LDA $X_2$ pocket. The LDA $X_2$ pocket has not been found
experimentally. In LDA$+$U (see the second and the third of Fig.\
\ref{FinX}), the corresponding band is pushed down below the
Fermi level and no $X_2$ pocket is present conforming to the
experiments. This is expected since correlation effects are more
important for slower electrons and the velocity near the pocket
is rather small. For the path with $J=0$, we find that the
removal of $X_2$ pocket is around $U=3\ eV$ far off from $U=4\
eV$ where the correct magnetic anisotropy is predicted. Notice
that the corresponding band is way below the Fermi level at $U=4\
eV$ (see the second of Fig.\ \ref{FinX}). For the path with the
experimental magnetic moments, we also find
that the removal of the $X_2$ point is near the point $U=1.9\ eV$ and $%
J=1.2\ eV$ where the correct magnetic anisotropy is predicted. A
very important point to notice is that the band that makes $X_2$
pocket in LDA is just below the Fermi level (see the third of
Fig.\ \ref{FinX}). This brings a suspicion that the point $U=1.9\
eV$ and $J=1.2\ eV$ is where the $X_2$ pocket just disappear. For
this reason we study the $X_2$ pockets at $U=1.7\ eV$ and $J=1.1\
eV$ (see the bottom of Fig.\ \ref{FinX}). The corresponding band
is just above the Fermi level forming a tiny pocket . This removal
of $X_2$ pocket near the point where the correct magnetic
anisotropy is predicted, strengthens the connection between MAE
and the absence of $X_2$ pocket.

\begin{figure}[htb]
\epsfig{file=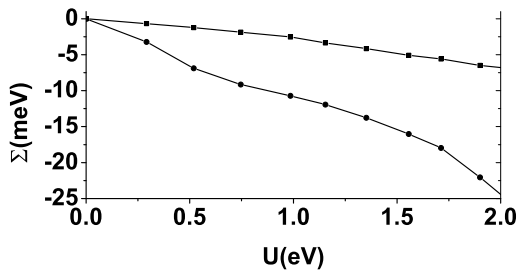, width=0.4\textwidth}
\vspace{6pt}
\caption{Diagonal Components of Self energy matrix
$\Sigma_{aa}^{\downarrow\downarrow}$ for minority spin.
Notice that $e_g$ band (circle) is more pushed down
than $t_{2g}$ band (square)}
\label{sigma}
\end{figure}

This disappearance of $X_2$ pocket can be explained in terms of
self energy $\Sigma_{ab}^{\sigma\sigma'}$. For simplicity, we
consider only diagonal components. Due to the cubic symmetry,
these diagonal components are the same for orbitals in $t_{2g}$
band and for orbitals in $e_g$ band:
\begin{equation}
\Sigma_{aa}^{\sigma\sigma}=\left\{
\begin{array}{cr}
\Sigma_{t_{2g}}^\sigma & \mbox{for orbitals in $t_{2g}$ bands} \\
\Sigma_{e_g}^\sigma & \mbox{for orbitals in $e_g$ bands}
\end{array}\right.
\end{equation}
  Since both $X_2$ pocket and
$X_5$ bands are dominated by minority spins, we consider only
minority spins here (see Fig.\ \ref{sigma}). At $U=1.9\ eV$ and
$J=1.2\ eV$, we find that $\Sigma_{t_{2g}}=-0.09 \ eV$ and
$\Sigma_{e_g}=-0.30\ eV$. This shows that $e_g$ bands are
suppressed more by strong correlation than $t_{2g}$ bands. Since
$X_2$ pocket is dominated by $e_g$ bands and $X_5$ pocket is
dominated by $t_{2g}$ bands, the $X_2$ pocket is removed by Strong
correlation effects while $X_5$ pocket survives. For comparison,
we find that $\Sigma_{t_{2g}}= -0.04\ eV$ and $\Sigma_{e_g}=-0.15\
eV$ at $U=1.1\ eV$ and $J=0.64\ eV$

There has been some suspicions that the incorrect position of the
$X_2$ band within LDA was responsible for the incorrect prediction
of the easy axis within this theory. Daalderop and
coworkers~\cite{dks} removed the $X_2$ pocket by increasing the
number of valence electrons and found the correct easy axis. We
therefore conclude that the absence of the pocket is one of the
central elements in determining the magnetic anisotropy, and there
is no need for any ad-hoc adjustment within a theory which takes
into account the correlations.

We now describe the effects originated from (near) degenerate
states close to the Fermi surface. These have been of primary
interest in past analytic studies~\cite{Kondorskii,mori:74}. We
will call such states {\em degenerate Fermi surface crossing}
(DFSC) states. The contribution to MAE by non-DFSC states comes
from the fourth order perturbation. Hence it is of the order of
$\lambda^4$, where $\lambda$ is spin--orbit coupling constant. The
energy splitting between DFSC states due to spin-orbit coupling
is of the order of $\lambda$ because the contribution comes from
the first order perturbation. Using linear approximation of the
dispersion relation $\epsilon(\vec{k}\lambda)$, the relevant
volume in $k$-space was found of the order $\lambda^3$. Thus,
these DFSC states make contribution of the order of $\lambda^4$.
Moreover, there may be accidentally near DFSC states appearing
along a line on the Fermi surface, rather than at a point. We
have found this to be the case in our LDA calculation for Ni.
Therefore the contribution of DFSC states is as important as the
bulk non-DFSC states though the degeneracies occur only in small
portion of the Brillouine zone.

This importance of the DFSC states leads us to comparative
analysis of the LDA and LDA+U band structures near the Fermi
level. In LDA (see top of Fig.\ \ref{FinX}), five bands are
crossing the Fermi level at nearly the same points along the
$\Gamma X$ direction. Two of the five bands are degenerate for the
residual symmetry and the other three bands accidentally cross the
Fermi surface at nearly the same points. There are two $sp$ bands
with spin up and spin down, respectively. The other three bands
are dominated by $d$ orbitals. In LDA$+$U (see second and third of
Fig.\ \ref{FinX}), one of the $d$ bands is pushed down below the
Fermi surface. The other four bands are divided into two
degenerate pieces at the Fermi level (see Fig. \ref{FinX}): Two
symmetry related degenerate $d_\downarrow$ bands and two near
degenerate $sp_\uparrow$ and $sp_\downarrow$ bands. In sum the
correlation weakens the effect of degenerate bands along $\Gamma
X$ direction.

In LDA (see the top of Fig.\ \ref{FinX}), we found that two bands
are accidentally near degenerate along the line on the Fermi
surface within the plane $\Gamma X L$. One band is dominated by
$d_\downarrow$ orbitals. The other is dominated by $d_\downarrow$
orbitals near $X$ and by $s_\downarrow$ orbitals off $X$. This
accidental DFSC states persist from $\Gamma X$ direction to
$\Gamma L$ direction. (Along $\Gamma L$ direction, the degeneracy
is for the residual symmetry). In LDA+U, these accidental DFSC
states disappear(see  Fig. \ref{fermi1}).   With the correlation
effect $U$, this accidental DFCS states along a line on the Fermi
surface move away from the Fermi surface leaving only the states
along $\Gamma L$ direction DFSC. This degenerate states' moving
away from Fermi surface makes the first order perturbation effect
sums up to zero.

\begin{figure}[htb]
\centering
        \caption{The L neck of LDA$+$U at $U=1.9\ eV$ $J=1.2\ eV$.
        Notice the absence of $L$ pockets.
nspecting the figures, we see that
there is a L neck. The solid lines represent bands with dominating up-spin.
         The dotted lines represent bands with dominating
         down-spin.  The horizontal line is the Fermi level. The points are:
         $N=(
 29/64,29/64,19/32)$,
$O= (30/64,30/64,18/32)$, $P=(
 31/64,31/64,17/32$,
        $Q=(125/256,125/256,67/128)$,
        $R\hspace{-0.05in}=\hspace{-0.05in}(126/256,126/256,66/128)$,\hspace{-0.05in}
 $S\hspace{-0.05in}=\hspace{-0.05in}(127/256,127/256,65/128)$, $L=(128/256,128/256,64/128)$.} \label{llleck}
\end{figure}

We now discuss another feature of Fermi surface, that is the $L$
neck. The experimental $L$ neck is a spin-up dominated band. From
the Fig.\ \ref{llleck}, we see one spin-up dominated band is just
below the Fermi level in $\Gamma L$ direction. As we move away
from the $\Gamma L$ direction, we see that the band is surfacing
up above the Fermi level. This is the experimentally confirmed
$L$ neck. This $L$ neck can be found at both $(U,J)=(1.9\
eV,J=1.2\ eV)$ and $(U,J)=(4.0\ eV,J=0\ eV)$. For $(U,J)=(4.0\
eV,J=0\ eV)$, we find additional $L$ pockets that has not been
found in experiments.

Based on the tight--binding model, the importance of DFSC states
has been shown~\cite{Kondorskii,mori:74}. We see that strong
correlations reduce number of DFSC states in $\Gamma X$ direction
and remove the near degenerate states on $\Gamma X L$ plane. We
conclude that the change of DFSC states is another important
element that determines the easy axis of Ni. The correct
configuration is tantamount to the correct Fermi surface. We see
that it is possible to find the correct Fermi surface at $U=1.9\
eV$ and $J=1.2\ eV$ where the magnetic moment and magnetic
anisotropy can be correctly predicted. A point to note is that
the Fermi surface does not come out right for other points in
$(U,J)$ parameter space.

\begin{figure}[htb]
\epsfig{file=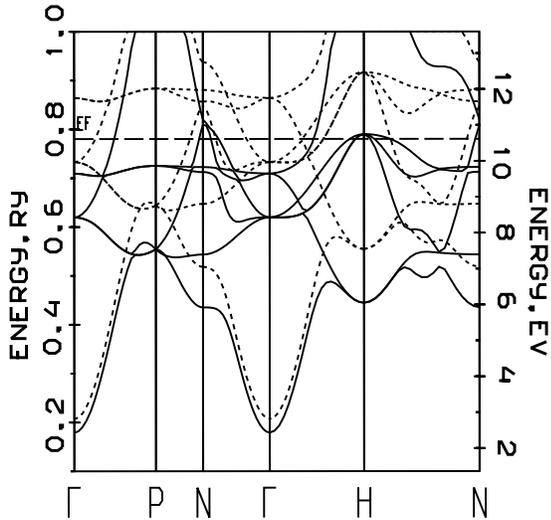,width=0.4\textwidth}
\vspace{6pt}
\caption{The band structure of Fe for LDA.
}
\label{fe_band}
\end{figure}

\begin{figure}[htb]
\epsfig{file=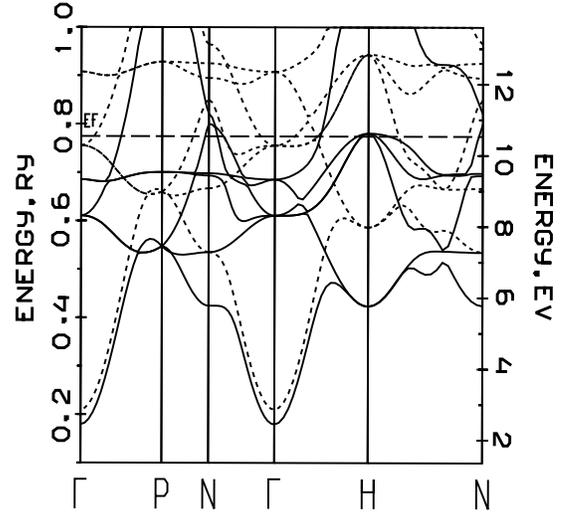,width=0.4\textwidth}
\vspace{6pt}
\caption{The band structure of Fe for   LDA$+$U at $U=1.2\ eV$ and $J=0.8\ eV$.
          }
\label{feu_band}
\end{figure}

Now we discuss band structure of Fe. Unlike Ni, strong correlation
effects do not change bands structure of Fe near the Fermi surface
significantly (see Fig.\ \ref{fe_band} and \ref{feu_band}). Most
of the minority spin $d$ bands lie above the Fermi surface and the
self energy is not large enough to push the bands down near the
Fermi surface. This is very different situation from that of Ni,
where minority spin $d$ bands are near the Fermi surface and the
self energy is of comparable magnitude. The self energy of
minority spin $e_g$ band is positive for Fe. Since most of this
band lies above the Fermi surface, the effects of strong
correlation to occupied states is insignificant. Majority spin
$d$ bands lie far below the Fermi level. Since the self energy
for this band is negative, the bands are pushed down more by
strong correlations. This is the same situation as that of Ni and
there is no significant changes near the Fermi level by strong
correlations. The band structure of LDA and LDA$+$U are drawn in
Fig.\ \ref{fe_band} and \ref{feu_band} . Comparing the two band
structure, one can see that there is no significant change near
the Fermi surface. This explains why magnetic anisotropy of Fe
has been solved in LSDA while magnetic anisotropy of Ni needs a
proper treatment of strong correlations.

\section{Spin--Other--Orbit Coupling}
\label{soosec}

Our implementation of relativistic effects neglects other
relativistic corrections than spin--orbit coupling. In this
section, we first present that other relativistic effect can be
seamlessly incorporated into out approach, then discuss
spin--other--orbit coupling, which might be important in
calculating magnetic anisotropy energy.

We can divide relativistic corrections into one-body interactions
and two-body interactions. One-body interactions are spin-orbit
coupling, relativistic mass correction, and Darwin term. These
one-body interactions can be included into Kohn--Sham Hamiltonian
in the same way as spin--orbit coupling is included. Only
spin-orbit coupling terms is relevant to magnetic anisotropy
energy and it has been the main theme of this paper.

There are two two-body interactions coming from relativistic
corrections, spin--other--orbit coupling and spin--spin
interaction:
\begin{eqnarray}
H_{SOO}&=&-\frac{\mu_B^{2}}{2}\sum_{i<j}\frac{1}{r_{ij}^{3}}\hat\R_{ij}\times
{\bf p}_{i}\cdot({\bf s}_{i}+2{\bf s}_{j}), \\
H_{SS}&=&\frac{\mu_B^{2}}{2}\sum_{i<j}\frac{1}{r_{ij}^{3}}%
[{\bf s}_{i}\cdot {\bf s}_{j}-(\R_{ij}\cdot \frac{ {\bf
s}_{i})({\bf r}_{ij}\cdot{\bf s}_{j})}{r_{ij}^{2}}].
\end{eqnarray}
We now present the scheme of including two body interactions. A
fermionic field operator $\Psi$ can be expanded along orthonormal
complete bases $\{\Phi_\alpha\}$ where $\alpha$ is a combined
index:
\begin{equation}
\Psi=\sum_\alpha a_\alpha \Phi_\alpha.
\end{equation}
In this bases, the canonical anti--commutator relation can be
written down in creation and annihilation operators:
$\{a_\alpha^\dagger, a_\beta\}=\delta_{\alpha\beta}$. In second
quantization approach, any two particle interaction can be written
down as
\begin{equation}
H_{\mbox{\scriptsize
int}}=\frac{1}{2}\sum_{\alpha\beta\gamma\delta} a^\dagger_\alpha
a^\dagger_\gamma a_\delta a_\beta U^{\alpha\beta\gamma\delta},
\end{equation}
 The matrix element
$U_{\alpha\beta\gamma\delta}$ can be found by
\begin{equation}
U_{\alpha\beta\gamma\delta} = \int\Phi _{\alpha}^{\dagger }({\bf
r}_1) \Phi_{\gamma}^{\dagger }({\bf r}_2)v(1,2) \Phi
_{\delta}({\bf r}_2)\Phi _\beta({\bf r}_1) d{\bf r}_1d{\bf r}_2,
\end{equation}
where $v(1,2)$ is a two particle interaction between particle $1$
and particle $2$. The expectation value of $H_{\mbox{\scriptsize
int}}$ can expressed as
\begin{equation}
\langle H_{\mbox{\scriptsize int}}
\rangle=\sum_{\alpha\beta\gamma\delta}
(n_{\alpha\beta}n_{\gamma\delta}-n_{\alpha\delta}n_{\gamma\beta})
U^{\alpha\beta\gamma\delta}. \label{aveint}
\end{equation}
Introducing another interaction matrix
\begin{equation}
J_{\alpha\beta\gamma\delta} = \int\Phi _{\alpha}^{\dagger }({\bf
r}_1) \Phi_{\gamma}^{\dagger }({\bf r}_2)v(1,2) \Phi _{\beta}({\bf
r}_2)\Phi _\delta({\bf r}_1) d{\bf r}_1d{\bf r}_2,
\end{equation}
The Eq.\ \ref{aveint} can be written as
\begin{equation}
\langle H_{\mbox{\scriptsize int}}
\rangle=\sum_{\alpha\beta\gamma\delta}
(U^{\alpha\beta\gamma\delta}-J^{\alpha\beta\gamma\delta})n_{\alpha\beta}n_{\gamma\delta}
. \label{aveint1}
\end{equation}

The approach described in section~\ref{ldaudmftsec} can be
straightforwardly extended to accommodate general two particle
interactions by allowing spin dependent interaction matrices
$U^{\alpha\beta\gamma\delta}$ and $J^{\alpha\beta\gamma\delta}$.
Therefore, all the relativistic corrections can be included in the
framework of DMFT and  its static limit LDA+U.

The Coulomb interaction we have considered is one of the two
particle interaction. In case of Coulomb interaction the
interaction matrix $U_{\alpha\beta\gamma\beta}$ is independent of
spin. For more rigorous discussion we decompose the index $\alpha$
to $(a,\sigma)$, where $\sigma$ is spin index and $a$ contains all
the others. The interaction matrix of Coulomb interaction is
\begin{eqnarray}
U_{\alpha\beta\gamma\beta} &=& U_{abcd}\delta_{\sigma_a\sigma_b}
\delta_{\sigma_c\sigma_d} \\ J_{\alpha\beta\gamma\beta} &=&
J_{abcd}\delta_{\sigma_a\sigma_d} \delta_{\sigma_c\sigma_b}.
  \label{imd}
\end{eqnarray}
The $U_{abcd}$ and $J_{abcd}$ interaction matrices are the same
interaction matrices in Eq.\ \ref{Emodel}.

 For general two particle
interaction, the self energy is for spin-diagonal elements
\begin{eqnarray}
\Sigma _{ab}^{\sigma \sigma
}&=&\sum_{cd}(U_{abcd}^{\sigma\sigma-\sigma-\sigma}
-J_{abcd}^{\sigma\sigma-\sigma-\sigma})n_{cd}^{-\sigma -\sigma }
\\ &+&\sum_{cd}(U_{abcd}^{\sigma\sigma\sigma\sigma}
-J_{abcd}^{\sigma\sigma\sigma\sigma})n_{cd}^{\sigma \sigma }
\nonumber  \\ &+&\sum_{cd}(U_{abcd}^{\sigma\sigma\sigma-\sigma}
-J_{abcd}^{\sigma\sigma\sigma-\sigma})n_{cd}^{\sigma- \sigma }
\nonumber  \\ &+&\sum_{cd}(U_{abcd}^{\sigma\sigma-\sigma\sigma}
-J_{abcd}^{\sigma\sigma-\sigma\sigma})n_{cd}^{-\sigma \sigma }
\nonumber\\ &-&\frac {%
\delta \Phi _{DC}}{\delta G_{ab}^{\sigma\sigma}(i\omega
)}.\nonumber\label{lambdasigmasoo}
\end{eqnarray}
Unlike the case of Coulomb interaction, the spin--diagonal
elements of self energy can get contributions from
spin--off-diagonal density matrix components. Notice that there
are contributions from spin-diagonal elements with opposite spin
through exchange interaction
($J_{abcd}^{\sigma\sigma-\sigma-\sigma}n_{cd}^{-\sigma -\sigma}
$). The spin off--diagonal elements of self energy $\Sigma
_{ab}^{\sigma -\sigma }$ takes the same form as the spin-diagonal
elements except that the spin corresponding to the index $b$ is
$-\sigma$ instead of $\sigma$.

We now study spin--other--orbit coupling. In this interaction, the
spin degrees of freedom sees spatial anisotropy through orbital
degrees of freedom. Spin--other-orbit coupling, therefore, may be
important in magnetic anisotropy, though its contribution to MAE
might be small. Since each term in spin-other-orbit coupling
contains only one spin operator, either spins of $\alpha$ and
$\beta$ or spins of $\gamma$ and $\delta$ must be the same in Eq.\
\ref{aveint}. Using this observation, the spin-diagonal elements
of self energy is simplified to
\begin{eqnarray}
\Sigma _{ab}^{\sigma \sigma
}&=&\sum_{cd}(U_{abcd}^{\sigma\sigma-\sigma-\sigma}
 )n_{cd}^{-\sigma
-\sigma } \\ &+&\sum_{cd}(U_{abcd}^{\sigma\sigma\sigma\sigma}
-J_{abcd}^{\sigma\sigma\sigma\sigma})n_{cd}^{\sigma \sigma }
\nonumber  \\ &+&\sum_{cd}(U_{abcd}^{\sigma\sigma\sigma-\sigma}
-J_{abcd}^{\sigma\sigma\sigma-\sigma})n_{cd}^{\sigma- \sigma }
\nonumber  \\ &+&\sum_{cd}(U_{abcd}^{\sigma\sigma-\sigma\sigma}
-J_{abcd}^{\sigma\sigma-\sigma\sigma})n_{cd}^{-\sigma \sigma }
\nonumber\\ &-&\frac {%
\delta \Phi _{DC}}{\delta G_{ab}^{\sigma\sigma}(i\omega )}
\nonumber ,\label{lambdasigma1}.
\end{eqnarray}
Notice that the contribution from spin-diagonal density matrix
components has the same form as the one of Coulomb interaction
except that $U_{abcd}^{\sigma\sigma\sigma\sigma}$ need not to be
the same as $U_{abcd}^{\sigma\sigma-\sigma-\sigma}$. Since the
spin-off-diagonal density matrix elements are small compared to
the spin-diagonal elements, we may neglect such contributions.
When Coulomb interaction is consider as well as spin--other
interaction, its contribution to the interaction matrix elements
of the form $U_{abcd}^{\sigma\sigma\sigma\sigma}$,
$U_{abcd}^{\sigma\sigma-\sigma-\sigma}$, or
$J_{abcd}^{\sigma\sigma\sigma\sigma}$ is much larger than that of
spin--other--coupling.
 In this
approximation, the contribution of spin--other--orbit coupling is
negligible and the spin-diagonal elements of self energy is then
the same as that of Coulomb interaction.

Spin off--diagonal elements is given by \begin{eqnarray} \Sigma
_{ab}^{\sigma -\sigma }&=&\sum_{cd}(
-J_{abcd}^{\sigma-\sigma-\sigma\sigma})n_{cd}^{-\sigma \sigma }
\nonumber\\ &+&\sum_{cd}(U_{abcd}^{\sigma-\sigma-\sigma-\sigma}
-J_{abcd}^{\sigma-\sigma-\sigma-\sigma})n_{cd}^{-\sigma -\sigma }
\\ &+&\sum_{cd}(U_{abcd}^{\sigma-\sigma\sigma\sigma}
-J_{abcd}^{\sigma-\sigma\sigma\sigma})n_{cd}^{\sigma \sigma }
\nonumber   \\ &-&\frac {%
\delta \Phi _{DC}}{\delta G_{ab}^{\sigma-\sigma}(i\omega
)}.\nonumber\label{lambdasigma2}
\end{eqnarray}
The contributions from spin--off--diagonal density matrix elements
has the same from as that of Coulomb interaction. Notice that
there are contributions from spin--diagonal density matrix
elements. This contribution should be treated on the same footing
as the contributions from spin--off--diagonal density matrix
elements, hence the correction from spin--other--orbit coupling is
not negligible. For an extensive study of this effect, we need an
implementable formula. We developed an formula that can be
implemented in Appendix

An extensive study on the effects of spin--other--coupling to
spin--off--diagonal elements of self energy requires another
paper. Instead, we estimate the order of magnitude of
spin--other--orbit coupling here. The order of magnitude of
spin--other--coupling can be estimated by using $\langle
1/r_{ij}^3 \rangle \sim \langle 1/r_{ij} \rangle^3$ to
$(CI/2\alpha m c^2)^2 CI$, where $CI$ is the strength of Coulomb
interaction and $\alpha$ is the fine structure constant. We see
that the order of magnitude of $CI$ is $1\ eV$ for $d$ shell
electrons of Ni and Fe. Therefore, the spin--other--interaction is
$10^{-8}$ times smaller than the Coulomb interaction. The effect
of spin--other--orbit coupling is, therefore negligible. This has
been confirmed by the study of Stiles and
coworkers~\cite{halilov1}.

\section{Conclusion} \label{concsec} We have studied the
effects of strong correlations on magnetic anisotropy. We find
that magnetic anisotropy   changes as correlation strength $U$ and
$J$ change.   These has been shown along several paths in $(U,J)$
space. The correct magnetic anisotropy can be predicted at
several points in $(U,J)$ space for both Ni and Fe. Moreover,
magnetic moment, magnetic anisotropy, and Fermi surface are
correctly predicted simultaneously at physically acceptable values
of $U$ and $J$: $U=1.9\ eV$ and $J=1.2\ eV$ for Ni and $U=1.2\ eV$
and $J=0.8\ eV$ for Fe.

It is remarkable that the values of $U$ necessary to reproduce the
correct magnetic anisotropy energy are very close (within 1.2 eV)
to the values which are needed to  describe photoemission spectra
of these materials~\cite{kl99,kl00}. The correct estimation of $U$
is indeed a serious problem. Different estimates based on
different methods give a range of values from $1\ eV$ to $6\ eV$.
The work of Katsnelson and coworkers used a dynamic approach
utilizing finite temperature Quantum Monte Carlo method whereas we
use a static approach at zero temperature. Considering these
discrepancies, an agreement within $1.2 eV$ shows an internal
consistency of our approach and emphasizes the importance of
correlations.

For Ni, we find that the LDA $X_2$ pocket disappears near $U=1.9\
eV$ and $J=1.2\ eV$ where the correct magnetic moment and the
correct magnetic anisotropy are predicted simultaneously. This
suggests a tight relation between the absence of $X_2$ pocket and
the correct magnetic anisotropy as suggested by previous work. The
change of DFSC states from LDA to LDA$+$U is discussed. DFSC
states are old suspects of dominant contribution to magnetic
anisotropy. The correct configuration of DFSC states and the
absence of $X_2$ pocket are by-products of the correct Fermi
surface that we predict for the first time. For Fe, we find that
strong correlations do not change band structures significantly.
This is attributed to the fact that most of minority bands lie
above the Fermi level, hence the change of band structures has no
large effect to occupied states. This distinguishes the magnetic
anisotropy problem for Fe from the one for Ni.

The calculations performed are state of the art in what can
currently be achieved for realistic treatments of correlated
solids.  Despite the great successes of the LDA+U theory in
predicting material properties of correlated solids (for a
review, see book of Anisimov \cite{anisimov}), there are obvious
problems of this approach when applied to metals or to systems
where the orbital symmetries are not broken. The most noticeable
is that it only describes spectra which has Hubbard bands. A
correct treatment of the electronic structure of strongly
correlated electron systems has to treat both Hubbard bands and
quasiparticle bands on the same footing. Another problem occurs in
the paramagnetic phase of Mott insulators, in the absence of any
broken symmetry the LDA + U method reduces to the LDA, and the gap
collapses. In systems like NiO where the gap is of the order of
eV, but the Neel temperature is a few hundred Kelvin, it is
unphysical to assume that the gap and the magnetic ordering are
related. For this reason the LDA+U predicts magnetic order in
cases that it is not observed, as, e.g., in the case of Pu
\cite{Pu-LDA+U}. Since LDA$+$U method is a limiting case of more
general DMFT, these difficulties are expected to be overcome in
DMFT. Further studies should be devoted to improving the quality
of the solution of the impurity model within DMFT and extending
the calculation to finite temperatures.

\acknowledgments

This research was supported by the ONR grants No.\@ 4-2650 and
N00014-02-1-0766. GK
would like to thank K. Hathaway for discussing the origin of
magnetic anisotropy and G. Lonzarich for discussing dHvA data. We
thank R. Chitra for stimulating discussions. We thank V. Oudovenko
and C. Uebing for developing and setting up the Beowulf 
computer cluster used to perform these
calculations. We have also used the supercomputer at the Center
for Advanced Information Processing, Rutgers.

\appendix
\section*{Appendix}
In this appendix, we develop an computationally implementable
formula for spin--other--orbit coupling:
\begin{equation}
v(1,2)=i\mu_B^2\frac{(\R_1-\R_2)\times \nabla_1}{r_{12}^3}
\cdot(\sigma_1+2\sigma_2)+1\leftrightarrow 2,
\end{equation}
where $\sigma_1$ applies to states at $\R_1$ and $\sigma_2$ to
states at $\R_2$. For simplicity, we will only work the first term
out. The formula for the second term can be found by exchanging
$1$ and $2$. Using the identity
\begin{equation}
\nabla=\hat\R\frac{\partial}{\partial r}-i\frac{\R\times \vec
L}{r^2},
\end{equation}
the spin-other-interaction can be written down as
\begin{eqnarray}
v(1,2)&=&-\mu_B^2\frac{1}{r_{12}^3} \vec
L_1\cdot(\sigma_1+2\sigma_2)  \\&+&
i\mu_B^2r_2\frac{(\hat\R_1)\times \hat\R_2}{r_{12}^3}
\cdot(\sigma_1+2\sigma_2)\frac{\partial}{\partial r_1}\nonumber \\
&-& \mu_B^2\frac{r_2}{r_1}\frac{\hat\R_2\cdot \hat\R_1}{r_{12}^3}
\vec L_1\cdot(\sigma_1+2\sigma_2) \nonumber\\&+&
\mu_B^2\frac{r_2}{r_1}\frac{\hat\R_2\cdot \vec L_1}{r_{12}^3}
\hat\R_1\cdot(\sigma_1+2\sigma_2). \label{soodec} \nonumber
\end{eqnarray}
We will write the interaction matrix as
\begin{equation}
U_{\alpha\beta\gamma\delta}=U^1_{\alpha\beta\gamma\delta}
+U^2_{\alpha\beta\gamma\delta}+U^3_{\alpha\beta\gamma\delta},
\end{equation}
where $U^1_{\alpha\beta\gamma\delta}$ comes from the first term,
$U^2_{\alpha\beta\gamma\delta}$ from the second term, and
$U^3_{\alpha\beta\gamma\delta}$ from the third and fourth terms of
the Eq.\ \ref{soodec}. For easiness of following index structure,
we define the interaction matrix $U$ as
\begin{equation}
U_{\alpha\beta\gamma\delta} = \int\Phi _{\alpha}^{\dagger }({\bf
r}_1) \Phi_{\beta}^{\dagger }({\bf r}_2)v(1,2) \Phi _{\gamma}({\bf
r}_2)\Phi _\delta({\bf r}_1) d{\bf r}_1d{\bf r}_2.
\end{equation}
We will consider bases in the from of
\begin{equation}
\Phi_{ilm\sigma}({\bf
r})=\phi_{il}(r)Y_{lm}(\theta,\phi)\chi_{\sigma},
\end{equation}
where $\chi_\sigma=(1,0)^T$, or $(0,1)^T$. Note that the angles
$\theta$ and $\phi$ is measured with respect to the origin of a
coordinate system.

We want to separate the spin-other-orbit interaction into radial
part and angular part. This can be done by using the following
property
\begin{equation}
\frac{1}{r_{12}^3}=\sum_{lm}
f_l(r_1,r_2)Y^*_{lm}(\theta_2,\phi_2)Y_{lm}(\theta_1,\phi_1),
\end{equation}
where
\begin{equation}
f_l(r_1,r_2)=4\pi\int^1_{-1}\frac{P_l(x)}{(r_1^2+r_2^2-2r_1r_2x)^{2/3}}dx.
\end{equation}
For example we write down $f_0$ and $f_1$ explicitly:
\begin{equation}
f_0(r_1,r_2)=\frac{4\pi}{r_1r_2}[\frac{1}{|r_1-r_2|}-\frac{1}{r_1+r_2}]
\end{equation}
\begin{eqnarray}
f_1(r_1,r_2)=\frac{2\pi}{r_1^2r_2^2}[&&\frac{(r_1-r_2)^2+r_1^2+r_2^2}{|r_1-r_2|}
\nonumber \\ &&-\frac{(r_1+r_2)^2+r_1^2+r_2^2}{r_1+r_2}].
\end{eqnarray}

With these identities, we can calculate a readily implementable
formula for spin--other--orbit interaction. After some
manipulation, we can write the contribution from the first terms
of the Eq.\ \ref{soodec} as
\begin{eqnarray}
&&U^1_{(i_1l_1m_1\sigma_1)(i_2l_2m_2\sigma_2)(i_3l_3m_3\sigma_3)(i_4l_4m_4\sigma_4)}\nonumber
\\ =&& -\mu_B^2\sum_{lm}\int dr_1 dr_2 f_l(r_1,r_2)\times\nonumber \\
&&\phi_{i_1l_1}^*(r_1)\phi_{i_2l_2}^*(r_1)\phi_{i_3l_3}(r_1)\phi_{i_4l_4}(r_1)\times
\nonumber\\ &&\langle l_2m_2|Y_{lm}| l_3m_3\rangle\langle
l_1m_1|Y_{lm}O_{\sigma_1\sigma_2\sigma_3\sigma_4}| l_4m_4\rangle,
\end{eqnarray} where the operator $O_{\sigma_1\sigma_2\sigma_3\sigma_4}^1$ is defined by\begin{equation}
O_{\sigma_1\sigma_2\sigma_3\sigma_4}^1=2\delta_{\sigma_1\sigma_4}O^1_{\sigma_2\sigma_3}+
\delta_{\sigma_2\sigma_3}O^1_{\sigma_1\sigma_4}
\label{soooone}\end{equation}
\begin{eqnarray}
O^1_{\uparrow\uparrow}&=&L_z\nonumber \\
O^1_{\downarrow\downarrow}&=&-L_z \nonumber\\
O^1_{\uparrow\downarrow}&=&L_- \nonumber\\
O^1_{\downarrow\uparrow}&=&L_+\nonumber
\end{eqnarray}

The contribution from the second terms of the Eq.\ \ref{soodec}
reads
\begin{eqnarray}
&&U^2_{(i_1l_1m_1\sigma_1)(i_2l_2m_2\sigma_2)(i_3l_3m_3\sigma_3)(i_4l_4m_4\sigma_4)}=\nonumber
\\
&&i\mu_B^2\sum_{lm}\int dr_1 dr_2 r_2 f_l(r_1,r_2)\times
\nonumber\\
&&\phi_{i_1l_1}^*(r_1)\phi_{i_2l_2}^*(r_1)\phi_{i_3l_3}(r_1)\frac{\partial}{\partial
r_1}\phi_{i_4l_4}(r_1)\times\nonumber
\\
&&O^2_{(l_1m_1\sigma_1)(l_2m_2\sigma_2)(l_3m_3\sigma_3)(l_4m_4\sigma_4)}(lm),
\end{eqnarray}
where
\begin{eqnarray}
&&O^2_{(l_1m_1\sigma_1)(l_2m_2\sigma_2)(l_3m_3\sigma_3)(l_4m_4\sigma_4)}(lm)\nonumber\\
&&=2\delta_{\sigma_1\sigma_4}N^{\sigma_2\sigma_3}_{(l_1m_1)(l_2m_2)(l_3m_3)(l_4m_4)}(lm)
\nonumber\\
&&+\delta_{\sigma_2\sigma_3}N^{\sigma_1\sigma_4}_{(l_1m_1)(l_2m_2)(l_3m_3)(l_4m_4)}(lm),
\end{eqnarray}
and the operator
$N^{\sigma_1\sigma_2}_{(l_1m_1)(l_2m_2)(l_3m_3)(l_4m_4)}(lm)$ is
defined as
\begin{eqnarray}
&&N^{
\uparrow\uparrow}_{(l_1m_1)(l_2m_2)(l_3m_3)(l_4m_4)}(lm)\nonumber\\&&=
A_{(l_1m_1)(l_2m_2)(l_3m_3)(l_4m_4)}^z(lm)\\ &&N^{
\downarrow\downarrow}_{(l_1m_1)(l_2m_2)(l_3m_3)(l_4m_4)}(lm)\nonumber\\&&=
-A_{(l_1m_1)(l_2m_2)(l_3m_3)(l_4m_4)}^z(lm)\\ &&N^{
\uparrow\downarrow}_{(l_1m_1)(l_2m_2)(l_3m_3)(l_4m_4)}(lm)\nonumber\\&&=
A_{(l_1m_1)(l_2m_2)(l_3m_3)(l_4m_4)}^-(lm) \\ &&N^{
\downarrow\uparrow}_{(l_1m_1)(l_2m_2)(l_3m_3)(l_4m_4)}(lm)\nonumber\\&&=
A_{(l_1m_1)(l_2m_2)(l_3m_3)(l_4m_4)}^+(lm).
\end{eqnarray}
The operator $\vec A_{(l_1m_1)(l_2m_2)(l_3m_3)(l_4m_4)}(lm)$ is
defined by a cross product of another operator $\vec
B_{(l_1m_1)(l_2m_2)}$:
\begin{eqnarray}
&&\vec A_{(l_1m_1)(l_2m_2)(l_3m_3)(l_4m_4)}(lm) \nonumber\\
&&=\vec B_{(l_1m_1)(l_4m_4)}(lm)\times \vec
B_{(l_2m_2)(l_3m_3)}(lm),
\end{eqnarray}
where
\begin{eqnarray}
&& B^x_{lml'm'}(lm)=\sqrt{\frac{2\pi}{3}}\langle
lm|Y_{lm}(Y_{1-1}-Y_{11})| l'm'\rangle \\ &&
B^y_{lml'm'}(lm)=\sqrt{\frac{2\pi}{3}}\langle
lm|Y_{lm}(Y_{1-1}+Y_{11})| l'm'\rangle \\ &&
B^z_{lml'm'}(lm)=\sqrt{\frac{4\pi}{3}}\langle lm|Y_{lm}Y_{10}|
l'm'\rangle
\end{eqnarray}

The contribution from the last two terms of the Eq.\ \ref{soodec}
is
\begin{eqnarray}
&&U^3_{(i_1l_1m_1\sigma_1)(i_2l_2m_2\sigma_2)(i_3l_3m_3\sigma_3)(i_4l_4m_4\sigma_4)}=\nonumber
\\
&&\mu_B^2\sum_{lm}\int dr_1 dr_2 \frac{r_2}{r_1}
f_l(r_1,r_2)\times
\nonumber\\&&\phi_{i_1l_1}^*(r_1)\phi_{i_2l_2}^*(r_1)\phi_{i_3l_3}(r_1)
\phi_{i_4l_4}(r_1)\times\nonumber
\\
&&\left[\frac{4\pi}{3}O^3_{(l_1m_1\sigma_1)(l_2m_2\sigma_2)(l_3m_3\sigma_3)(l_4m_4\sigma_4)}(lm)\right.\nonumber
\\
&&\left.-\frac{2\pi}{3}O^4_{(l_1m_1\sigma_1)(l_2m_2\sigma_2)(l_3m_3\sigma_3)(l_4m_4\sigma_4)}(lm)\right].
\end{eqnarray}
The operator
$O^3_{(l_1m_1\sigma_1)(l_2m_2\sigma_2)(l_3m_3\sigma_3)(l_4m_4\sigma_4)}(lm)$,
which represents the contribution from the third term, is defined
by
\begin{eqnarray}
&&O^3_{(l_1m_1\sigma_1)(l_2m_2\sigma_2)(l_3m_3\sigma_3)(l_4m_4\sigma_4)}(lm)\nonumber
\\ &&=\langle l_2m_2|Y_{lm}Y_{10}| l_3m_3\rangle\times\nonumber\\&&\langle l_1m_1|Y_{lm}Y_{10}O^1_{\sigma_1\sigma_2\sigma_3\sigma_4}|
l_4m_4\rangle\nonumber
\\ &&-\langle l_2m_2|Y_{lm}Y_{11}| l_3m_3\rangle\times\nonumber\\&&\langle l_1m_1|Y_{lm}Y_{1-1}O^1_{\sigma_1\sigma_2\sigma_3\sigma_4}|
l_4m_4\rangle\nonumber
\\ &&-\langle l_2m_2|Y_{lm}Y_{1-1}| l_3m_3\rangle\times\nonumber\\&&\langle l_1m_1|Y_{lm}Y_{11}O^1_{\sigma_1\sigma_2\sigma_3\sigma_4}|
l_4m_4\rangle,
\end{eqnarray}
where $O^1_{\sigma_1\sigma_2\sigma_3\sigma_4}$ is given by the
Eq.\ \ref{soooone}. The operator
$O^4_{(l_1m_1\sigma_1)(l_2m_2\sigma_2)(l_3m_3\sigma_3)(l_4m_4\sigma_4)}(lm)$
which represents the contribution from the last term, is defined
by
\begin{eqnarray}
&&O^4_{(l_1m_1\sigma_1)(l_2m_2\sigma_2)(l_3m_3\sigma_3)(l_4m_4\sigma_4)}(lm)\nonumber
\\ &&=\sqrt{2}\langle l_2m_2|Y_{lm}Y_{10}| l_3m_3\rangle\times\nonumber\\&& \langle l_1m_1|Y_{lm}O^4_{\sigma_1\sigma_2\sigma_3\sigma_4}L_z|
l_4m_4\rangle\nonumber
\\ &&-\langle l_2m_2|Y_{lm}Y_{11}| l_3m_3\rangle\times\nonumber\\&&\langle l_1m_1|Y_{lm}O^4_{\sigma_1\sigma_2\sigma_3\sigma_4}L_-|
l_4m_4\rangle\nonumber
\\ &&-\langle l_2m_2|Y_{lm}Y_{1-1}| l_3m_3\rangle\times\nonumber\\&&\langle l_1m_1|Y_{lm}O^4_{\sigma_1\sigma_2\sigma_3\sigma_4}L_+|
l_4m_4\rangle,
\end{eqnarray}
where $O_{\sigma_1\sigma_2\sigma_3\sigma_4}^4$ is given by the
following equations:
\begin{equation}
O_{\sigma_1\sigma_2\sigma_3\sigma_4}^4=2\delta_{\sigma_1\sigma_4}O^4_{\sigma_2\sigma_3}+
\delta_{\sigma_2\sigma_3}O^4_{\sigma_1\sigma_4} \end{equation}
\begin{eqnarray}
O^1_{\uparrow\uparrow}&=&\sqrt{2} Y_{10} \nonumber\\
O^1_{\downarrow\downarrow}&=&-\sqrt{2} Y_{10}\nonumber\\
O^1_{\uparrow\downarrow}&=&Y_{1-1} \nonumber\\
O^1_{\downarrow\uparrow}&=&-Y_{11}.\nonumber
\end{eqnarray}

\bibliographystyle{prsty}

\end{document}